\documentclass[lettersize,journal]{IEEEtran}
\usepackage{amsmath,amsfonts}
\usepackage{algorithmic}
\usepackage{array}
\usepackage[caption=false,font=normalsize,labelfont=sf,textfont=sf]{subfig}
\usepackage{textcomp}
\usepackage{stfloats}
\usepackage{url}
\usepackage{verbatim}
\usepackage{graphicx}
\usepackage[utf8]{inputenc}
\usepackage{graphicx}
\usepackage{epstopdf}
\usepackage{multirow}
\usepackage{amsmath}
\usepackage{tabularx}
\usepackage{booktabs}
\usepackage{float}
\usepackage[linesnumbered,algoruled,boxed,lined]{algorithm2e}
\usepackage{array}
\usepackage{color}
\usepackage{amssymb}
\usepackage{cite}
\usepackage{soul}

\newtheorem{remark}{Remark}

\usepackage{adjustbox}
\usepackage{setspace}

\def\BibTeX{{\rm B\kern-.05em{\sc i\kern-.025em b}\kern-.08em
    T\kern-.1667em\lower.7ex\hbox{E}\kern-.125emX}}
\usepackage{balance}
\begin{document}
\title{Joint Power and 3D Trajectory Optimization for UAV-enabled Wireless Powered Communication Networks with Obstacles}
\author{\IEEEauthorblockN{Hongyang Pan, Yanheng Liu, Geng Sun,~\IEEEmembership{Member,~IEEE}, Junsong Fan, Shuang Liang, Chau Yuen,~\IEEEmembership{Fellow,~IEEE}}

 \thanks{This study is supported in part by the National Natural Science Foundation of China (61872158, 62002133, 62172186, 62272194), in part by the Science and Technology Development Plan Project of Jilin Province (20210101183JC, 20210201072GX), in part by the Young Science and Technology Talent Lift Project of Jilin Province (QT202013), and in part by China Scholarship Council. (\textit{Corresponding author: Geng Sun}).\protect}
	
\thanks{Hongyang Pan is with the College of Computer Science and Technology, Jilin University, Changchun 130012, China, and also with the Engineering Product Development (EPD) Pillar, Singapore University of Technology and Design, Singapore 487372 (e-mail: panhongyang18@foxmail.com). 

\par Yanheng Liu and Geng Sun are with the College of Computer Science and Technology, Jilin University, Changchun 130012, China, and also with the Key Laboratory of Symbolic Computation and Knowledge Engineering of Ministry of Education, Jilin University, Changchun 130012, China (e-mail: yhliu@jlu.edu.cn, sungeng@jlu.edu.cn).

\par Junsong Fan is with the College of Computer Science and Technology, Jilin University, Changchun 130012, China (e-mail: fanjs19@foxmail.com). 

\par Shuang Liang is with the School of Information Science and Technology, Northeast Normal University, Changchun, 130024, China (e-mail: liangshuang@nenu.edu.cn). 

\par Chau Yuen is with the Engineering Product Development (EPD) Pillar, Singapore University of Technology and Design, Singapore 487372 (e-mail: yuenchau@sutd.edu.sg). 

\par {This manuscript has been accepted by IEEE Transactions on Communications, doi: 10.1109/TCOMM.2023.3240697.}
}}
\markboth{Journal of \LaTeX\ Class Files,~Vol.~14, No.~8, August~2021}%
{Shell \MakeLowercase{\textit{et al.}}: A Sample Article Using IEEEtran.cls for IEEE Journals}

\maketitle

\begin{abstract}
	Unmanned aerial vehicle (UAV)-enabled wireless powered communication networks (WPCNs) are promising technologies in 5G/6G wireless communications, while there are several challenges about UAV power allocation and scheduling to enhance the energy utilization efficiency, considering the existence of obstacles. In this work, we consider a UAV-enabled WPCN scenario that a UAV needs to cover the ground wireless devices (WDs). During the coverage process, the UAV needs to collect data from the WDs and charge them simultaneously. To this end, we formulate a joint-UAV power and three-dimensional (3D) trajectory optimization problem (JUPTTOP) to simultaneously increase the total number of the covered WDs, increase the time efficiency, and reduce the total flying distance of UAV so as to improve the energy utilization efficiency in the network. Due to the difficulties and complexities, we decompose it into two sub optimization problems, which are the UAV power allocation optimization problem (UPAOP) and UAV 3D trajectory optimization problem (UTTOP), respectively. Then, we propose an improved non-dominated sorting genetic algorithm-II with $K$-means initialization operator and Variable dimension mechanism (NSGA-II-KV) for solving the UPAOP. For UTTOP, we first introduce a pretreatment method, and then use an improved particle swarm optimization with Normal distribution initialization, Genetic mechanism, Differential mechanism and Pursuit operator (PSO-NGDP) to deal with this sub optimization problem. Simulation results verify the effectiveness of the proposed strategies under different scales and settings of the networks. 
\end{abstract}

\begin{IEEEkeywords}
	Wireless powered communication networks, unmanned aerial vehicle, energy consumption, non-dominated sorting genetic algorithm-II, particle swarm optimization.
\end{IEEEkeywords}

\section{Introduction}
\label{Introduction}

\par Recently, energy harvesting has attracted attention as a promising solution to prolong the lifetime of wireless networks in both academia and industry. Energy harvesting can be divided into two categories. The first one is the energy from natural renewable sources, such as solar and tidal energy, which provides renewable energy for wireless networks. However, due to the intermittence of these natural renewable sources, the energy supplies of wireless networks are not always reliable. The second one is the energy from far-field radio frequency (RF) signals, which are more controllable and stable than the energy from natural renewable sources. Moreover, RF signal can be utilized for both wireless information transmission (WIT) and wireless power transfer (WPT) in different network configurations \cite{7349240}.

\par Wireless powered communication network (WPCN) is a technique which combines the advantages of WIT and WPT. In general, a WPCN consists of multiple wireless devices (WDs) that can collect energy, a base station (BS) and a charger. Specifically, the charger transfers energy for WDs in the downlink, and the BS collects data from WDs in the uplink. 

\par With the development of WPT, two types of wireless chargers are usually utilized in WPCNs. The first one is called energy access point (EAP), which is only responsible for providing energy, and the second one is called the hybrid access point (HAP), which can supply extra data communication. In other words, HAP is as a charger and a data collecting device simultaneously \cite{7337464}. Compared to EAP as a wireless charger, the HAP charger has attracted more interests since it can support almost full-duplex HAP, where the energy transmission and data transmission can use different frequency bands. Moreover, in order to further improve the spectral efficiency, an HAP can transmit energy and receive data of WDs simultaneously by adopting advanced self-interference cancelation techniques \cite{2014Optimal}. However, there are some challenges in WPCN. For example, it is uneconomical to use a fixed device for charging and data collection, since the WDs are widely spread. It may need plenty of chargers and BSs. Thus, using unmanned aerial vehicles (UAVs) as chargers and BSs is a widely used solution due to their high mobility \cite{yang2020offloading}. However, owing to the limited on-board energy and the requirement of timeliness, UAVs need to allocate power and be scheduled reasonably to improve the energy utilization efficiency.

\IEEEpubidadjcol
\par To improve the energy utilization efficiency between a UAV and WDs, it is vital to schedule the UAV and allocate the powers of the UAV and WDs designedly since it is better to serve more WDs within limited time. However, a UAV may suffer several challenges for achieving the abovementioned purpose. For example, there are always some obstacles in practical scenarios. Under these circumstances, the difficulty of scheduling a UAV has undoubtedly increased, while the UAV cannot collision obstacles. Moreover, where to deploy the UAV is a vital problem, since the deployment may influence the energy consumption of UAV. Furthermore, the power allocation issue will influence the time efficiency of UAV. In addition, the three-dimensional (3D) trajectory of UAV also needs to be considered since it may directly affect the motion energy consumption of UAV. 

\par In order to solve the abovementioned problems, we jointly consider the hovering points, the power allocation and the energy consumption of UAV for enhancing the energy utilization efficiency under the environment with obstacles. The main contributions of this work can be summarized as follows:

\begin{itemize}
	\item We consider a UAV-enabled WPCN system, where a UAV can either hover to cover all WDs within the maximum coverage range or fly to next hovering point so that enhancing the energy utilization efficiency of the system. During the coverage process, the UAV needs to collect data from the WDs one by one and charge them simultaneously, which means that the charging process and data collection process are concurrent. For this purpose, we formulate a joint-UAV power and 3D trajectory optimization problem (JUPTTOP) to simultaneously increase the total number of covered WDs, improve the time efficiency and reduce the total flying distance of UAV.
	\item We prove JUPTTOP as an NP-hard problem, which has a large solution search space and it is difficult to be solved optimally. Moreover, the dimension of JUPTTOP is not fixed. Thus, we decompose the formulated JUPTTOP into two sub optimization problems, which are UAV power allocation optimization problem (UPAOP) and UAV 3D trajectory optimization problem (UTTOP), respectively.
	\item For UPAOP, we propose an improved non-dominated sorting genetic algorithm-II with $K$-means initialization operator and Variable dimension mechanism (NSGA-II-KV) to solve it. For UTTOP, we introduce a pretreatment method, then use an improved particle swarm optimization with Normal distribution initialization, Genetic mechanism, Differential mechanism and Pursuit operator (PSO-NGDP) to solve this problem.
	\item Through simulation, we verify that the NSGA-II-KV can effectively solve the converted UPAOP compared with the corresponding comparison algorithms. We can increase the coverage by $18.03\%$ at most while increasing the time efficiency by $2.93\%$. Moreover, the proposed PSO-NGDP is effective for the converted UTTOP. The flight energy consumption is reduced by $25.30\%$ at most. Such saving can be increased when the area of coverage is increased.
\end{itemize}

\par The rest of this paper is organized as follows. Section \ref{Related work} introduces the related work. Section \ref{System model} shows the system model. Section \ref{Problem statement} formulates the JUPTTOP. Section \ref{Algorithm for UPAOP} and Section \ref{Algorithm for UTTOP}  propose the algorithms for solving UPAOP and UTTOP, respectively. Section \ref{Simulation results} shows the simulation results. Section \ref{Discussion} discusses the influence of propulsion power of UAV and the rationality of UAV altitude. Finally, Section \ref{Conclusion} concludes this paper.

\section{Related work}
\label{Related work}

\par As a charger of a BS, researchers usually hope it can charge or communicate with more WDs. Thus, several previous works are dedicated to extending the coverage rates of UAVs. For example, Li \emph{et al}. \cite{8758340} considered an urban UAV seamless network scenario, where multiple UAVs cooperated to improve the coverage rates of the emergency situations. Based on the background of forest fire early warning and monitoring, the authors in \cite{9105056} formulated a min-time max-coverage issue in sweep coverage. Then, they proposed a heuristic algorithm weighted targets sweep coverage to find the optimal trajectory of UAV. Alkama \emph{et al}. \cite{9678360} proposed an analytical framework to analyze the coverage probability and capacity by using stochastic geometry. Moreover, they further derived the downlink coverage expression based on signal-to-interference-plus-noise-ratio.

\par Many prior works have considered to adopt movable devices such as UAVs and movable vehicles instead of conventional fixed chargers/BSs to achieve WPT/WIT\cite{9254114}, and these works can be divided into two categories. Specifically, the first one is about the transmission protocol design to share the limited time-frequency resources for both multiuser uplink WIT of uplink and downlink WPT. For example, the authors in \cite{2018Throughput} considered a time-division multiple access transmission protocol in which they aimed to increase the uplink minimum throughput among all WDs over a finite UAV flight period. The second one is to consider the joint trajectory and resource allocation design in UAV-enabled WPCN. On parallel, Miao \emph{et al}. \cite{9248523} considered a UAV-enabled WPCN applied for millimeter wave, where the transmission power and energy transfer time of multiple UAVs were optimized to obtain much higher throughput. The authors in \cite{9484496} considered WPT and WIT separately, i.e., one UAV served as a charger and another one acted as a BS. Then, a novel multi-agent deep Q-network (DQN) was proposed to optimize the trajectories of these two UAVs. Moreover, a resource allocation problem in UAV-enabled WPCN was investigated in \cite{8086177}, where RF energy was adopted to ensure the device-to-device communication. Due to the complexity of the problem, an efficient resource allocation algorithm based on Lagrangian relaxation method was considered to solve the problem. 
\par Several prior works studied the trajectory of UAV in the scenario with obstacles. For example, Shao \emph{et al}. \cite{9635801} discretized the flying time of UAV as a set of Chebyshev points and proposed a novel particle swarm optimization (PSO) to make UAV avoiding obstacles. The authors in \cite{8943975} proposed a new method with three stages, which were the initial trajectory generation, the trajectory correction, and the smooth to accelerate the convergence of the ant colony optimization algorithm. Experimental results verified the effectiveness of UAV collision avoidance. Based on 3D velocity obstacle spherical cap, Yang \emph{et al}. \cite{8536788} provided a direct obstacle avoidance method in dynamic space, and the proposed method could re-plan the online obstacle avoidance trajectory by using insertion point. Moreover, a safe-DQN approach was proposed in \cite{9385412} to maximize the uplink throughput of UAV by optimizing the trajectory in post-disaster scenarios. Specifically, the UAV could make decisions to select the optimal action in reasonable policy sets and the simulation results verified the performance of obstacle avoidance.

\par In addition, there were some previous works which studied the power allocation of the UAV. For instance, in \cite{DBLP:journals/tvt/MamaghaniH21}, the authors proposed a novel cooperative secure UAV aided transmission protocol while employing an artificial noise. For this purpose, they formulated an average secrecy rate (ASR) maximization problem, wherein a UAV was employed to deliver confidential data to a ground destination in the presence of a terrestrial passive eavesdropper. However, the UAV energy consumption was ignored in their work. The authors in \cite{DBLP:journals/access/MamaghaniH19} proposed a protocol to find an optimal UAV trajectory and proper resource allocation in UAV-assisted wireless communication for maximizing ASR. The simulations demonstrated that it could preform better with the aid of destination cooperative jamming and simultaneous wireless information and power transfer (SWIPT) at the UAV-relay. However, they ignored the obstacles which can  influence the trajectory of UAV. Then they jointly optimized the trajectory of a UAV, network transmission power, and artificial noise power allocation to enhance the ASR in \cite{DBLP:journals/access/MamaghaniH21}. For this purpose, they formulated a novel ASR maximization problem, subject to some quality of service and mission time constraints. However, the UAV could not deal with the obstacles in 3D environments. Furthermore, based on Terahertz technology, they considered the secrecy energy efficiency in \cite{DBLP:journals/tvt/MamaghaniH22}. Specifically, they assumed that the UAV-mounted relay might play a relay and a potential eavesdropper simultaneously, while the UAV could collect data acquisition from multiple ground WDs. However, they only assumed that the UAV flew at a fixed altitude, which could limit the 3D mobility of UAV.

\section{System model}
\label{System model}

\par In this section, the network model, the wireless charging model, the wireless data collection model, the energy consumption model of UAV and the obstacle model are introduced, and Table \ref{table:notations} lists the main notations of the variables for the readability and consistency.
\begin{table*}[t]
	\scriptsize
	\centering
	\caption{List of important Notations used in this paper}
	\label{table:notations}
	\renewcommand\arraystretch{1.3}
	\begin{tabular}{l|l|l|l}\hline
		Variable  &Physical meanings &Variable&Physical meanings\\\hline
		
		$N$ &Number of WDs &$M$ &Number of hovering points \\\hline
		
		$S_i$ &$i$th WD &$q_j$ &$j$th hovering point\\\hline
		
		$d_{ij}$ & Distance between $S_i$ and $q_j$ &$Ac_j$ & Number of covered WDs when UAV at $q_j$ \\\hline
		
		$\mu_{ij}$ &Charging efficiency between $S_i$ and $q_j$ &$d_{max}$ & Maximum coverage range \\\hline
		
		$Pc^{1\times M}$ & Transmission power of UAV for charging &$Pt^{1\times A_{c}}$ &Transmission power of a WD for data collection\\\hline
		
		$Pr^{N\times M}$ &Received charging power of a WD &$R^{N\times M}$ & Achievable rate of UAV \\\hline

		$R_{th}$ & Minimum achievable rate threshold &$Q^{1\times M}$ & Set of coordinates of UAV hovering points \\\hline
		$Th^{1\times M}$ &Hovering time of UAV &$e$ & Required energy of a WD \\\hline
		
		$u$ &Size of collected data of a WD &$Td^{1\times M}$ & Time difference between charging and data collection \\\hline
		
		$n$ &Number of hover points that cannot cover any WD &$w_j^k$& $k$th waypoint between $q_j$ and $q_{j+1}$ \\\hline
		
		$K$ & Number of waypoints between two hovering points &$Q^{*{M\times K}}$& Set of coordinates of waypoints \\\hline
		$Ed_{j}^k$ & Distance between $q_j$ and $w_j^k$   &$Ed_{j}^{k, k+1}$ & Distance between $w_j^k$ and $w_j^{k+1}$ \\\hline
		
		$(x_m^l, y_m^l)$ & Horizontal center of the $m$th obstacle &$h_m$ & altitude parameter of $m$th obstacle \\\hline
		
		$x_m^s, y_m^s$ & Parameters for controlling slope of $m$th obstacle &$Ob(x,y)$ & altitude of obstacles at $(x,y)$\\\hline			
	\end{tabular}
\end{table*}%

\subsection{Network model}

\par
\begin{figure}[t]
	\centering{\includegraphics[width=3.5in]{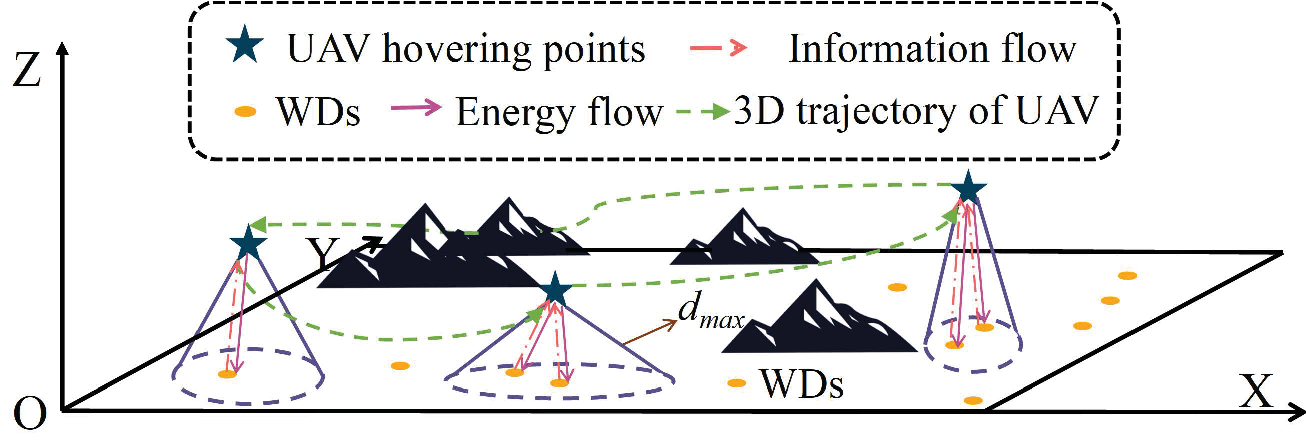}}
	\caption{A UAV-enabled WPCN with obstacles.}
	\label{Sys-model}
\end{figure}

 As shown in Fig. \ref{Sys-model}, we consider a UAV-enabled WPCN, which includes a UAV and $N$ WDs. Without loss of generality, the UAV can sail freely between the lowest altitude $Z_{min}$ and the highest altitude $Z_{max}$ at a constant speed and the WDs are at the ground, which can be denoted by $(S_{1}, S_{2}, S_{3}, ..., S_{N})$. Moreover, the 3D coordinates of the WDs are defined as ${S}_i = (X_i, Y_i, 0)$. During the flight, the collision with the obstacles should be avoided, and the UAV follows fly-hover-communication protocol mentioned in \cite{8663615}, i.e., the UAV needs to select a set of hovering points to cover WDs within the maximum coverage range. This coverage supports the UAV to charge all covered WDs within the maximum coverage range simultaneously and collect data from them one by one, which means that the charging and data collection process are concurrent. If the distance between the UAV and a WD is greater than the maximum coverage range, the UAV can neither charge the WD nor achieve the data collection with the WD. We define the 3D coordinates of the hovering points as ${q}_j = (x_j, y_j, z_j)$. Thus, the distance between ${S}_i$ and ${q}_j$ can be expressed as follows:
\begin{equation}
\label{d_{ij}}
\begin{aligned}
d_{ij}&=\sqrt{(X_i-x_j)^2+(Y_i-y_j)^2+{z_j}^2}
\end{aligned} 
\end{equation}

\subsection{Wireless charging model}

\par The purpose of charging is to extend the lifetime of WPCNs. Specifically, if a WD depletes its energy, it cannot work for sensing and communicating. However, it is inconvenient to replace batteries frequently for such WDs, especially in a large-scale WPCNs, since the WDs may be deployed in the forests or hills. As mentioned above, RF signal can be used for charging. A famous realization to use this method is radio frequency identification (RFID) \cite{2012Energy}, where passive RFID tags attached to objects may reply with an electronic product code-compliant ID when queried by a reader, by obtaining energy through the RF signal and reflecting the energy back. The paper-thin wireless identification and sensing platform (WISP) tags can harvest energy from RF signals transmitted by the readers, where WISP is of similar size as a 10-cent coin, and it can be easily attached to objects such as UAVs and WDs. Thus, we adopt RFID in \cite{8260857} to support energy for WDs due to its mentioned advantages. Referred to \cite{8260857}, the charging efficiency can be defined as follows:

\begin{equation}
\label{mu}
\begin{aligned}
{\mu}_{ij}=\left\{
\begin{matrix}
\frac{\gamma}{(d_{ij}+\tau)^2},&d_{ij} \leq d_{max}\\ 
0,&~\text{otherwise}
\end{matrix}\right.
\end{aligned}
\end{equation}

\noindent where $d_{max}$ is the maximum coverage distance. $\gamma$ is a constant related to the gains of the transmitting and receiver antennas, polarization loss, wavelength and rectifier efficiency. Moreover, $\tau$ is an adjust parameter of Friis' free space equation for the short distance transmission, which is also a constant. Thus, received power of the $i$th WD $S_i$ from $q_j$ is as follows:
\begin{equation}
\label{Pc_{ij}}
\begin{aligned}
Pr_{ij}={\mu}_{ij} \cdot Pc_{j}
\end{aligned}
\end{equation}
\noindent where $Pc_{j}$ is the transmission power of UAV for charging when UAV is at $q_j$. In this charging model, UAV can charge all covered WD simultaneously.
\begin{remark}
	\label{Chargingthreshold}
	{Note that there is a threshold value for charging. Generally, the WDs may be with the status of monitoring, sensing and communicating, etc \cite{DBLP:journals/ton/SrivastavaK13}. The energy threshold of a WD is to ensure that the WD can work normally for a period of time instead of fully charging the WD. The main reason is that if we fully charge the WD, it will extend the hovering time so as to increase the hovering energy consumption. Moreover, due to the periodicity of system tasks including sensing and communicating, we charge the WD to the threshold to ensure that they can work normally for a period of time. When the energy is almost exhausted, the next charging round will come, then the energy can continue to be replenished. Thus, it is unnecessary to fully charge the WD \cite{DBLP:journals/ton/SrivastavaK13}. Specifically, we assume that the considered WPCN in this work has run for a period of time, which means that all WDs have sufficient data while having almost no energy. Thus, the energy threshold is the required energy of a WD. After a period of time, WDs that almost deplete their own energy are those WDs executing sensing tasks more frequently, and such WDs will have sufficient data to collect by UAV once again. Then the UAV only needs to charge such WDs while collecting data from them in the next round flight. Compared to the considered WPCN in this work, such situation is simpler since the number of WDs may be reduced, thus our charging strategy is still feasible.}
\end{remark}

%

\subsection{Wireless data collection model}
\par During the hovering process, the UAV needs to collect data from the corresponding WDs. The data collection model is considered as a typical Air-to-Ground communication scenario. Let $h_{ij}$ denote the complex-valued channel coefficient between $S_i$ and ${q}_j$ and it can be expressed as follows \cite{8918497}:

\begin{equation}
\label{h_{ij}}
\begin{aligned}
h_{ij}=\sqrt{{\beta}_{ij}} \tilde{h}_{ij}
\end{aligned}
\end{equation}
\noindent where ${\beta}_{ij}$ accounts for the large-scale fading effects, and $\tilde {h}_{ij}$ accounts for the small-scale fading that is a complex random variable with $\mathbb{E}[{|\tilde{h}_{ij}|}^2]=1$. Without loss of generality, the elevation angle-dependent probabilistic line-of-sight (LoS) model is considered in this work, since UAV can obtain the probabilistic LoS link by moving, even in case of the presence of any obstacles. With this model, the large-scale fading is usually regarded as a random variable depending on the occurrence probabilities of LoS and non-LoS (NLoS) environments. Then, the large-scale channel coefficient ${\beta}_{ij}$ in Eq. (\ref{h_{ij}}) can be modeled as follows \cite{8918497}:
\begin{equation}
\label{beta_ij}
\begin{aligned}
{\beta}_{ij}=\left\{
\begin{matrix}
\beta_{0}{d_{ij}}^{-\alpha},&~\text{LoS environment}\\ 
\kappa \beta_{0}{d_{ij}}^{-\alpha},&~\text{NLoS environment}
\end{matrix}\right.
\end{aligned}
\end{equation}
\noindent where $\beta_{0}$ and $\alpha$ represent the path loss at the reference distance and the path loss exponent, respectively. $\kappa$ is the additional fading factor due to the NLoS environment. Moreover, the probability of having LoS environment is regarded as a logistic function of the elevation angle $\theta_{ij}$, which can be expressed as follows \cite{8918497}:
\begin{equation}
\label{P_LoS}
\begin{aligned}
P_{ij, LoS}(\theta_{ij})=\frac{1}{1+C\exp (-D[\theta_{ij}-C])}
\end{aligned}
\end{equation}
\noindent where $C$ and $D$ are modeling parameters, and $\theta_{ij}=\frac{180}{\pi}\sin^{-1}\left(\frac{z_j}{d_{ij}}\right)$ is the elevation
angle in degree. Moreover, the probability of NLoS is clearly given by $P_{ij, NLoS}=1-P_{ij, LoS}$. Thus, the expected channel power gain by averaging over
both randomness is as follows:
\begin{equation}
\label{E}
\begin{aligned}
\mathbb{E}[{|{h}_{ij}|}^2]=P_{ij, LoS}\beta_{0}{d_{ij}}^{-\alpha}+P_{ij, NLoS}\kappa \beta_{0}{d_{ij}}^{-\alpha}
\end{aligned}
\end{equation}
\noindent As mentioned above, we assume the UAV can collect data from the covered WDs one by one. Let $Pt_i$ represents the transmission power of the covered WD $S_i$, then the achievable rate can be expressed as follows \cite{8852863}:
\begin{equation}
\label{R_{ij}}
\begin{aligned}
R_{ij}=B\log_{2}\left(1+\frac{Pt_i{|{h}_{ij}|}^2}{\sigma^2}\right)
\end{aligned}
\end{equation}
\noindent where $B$ and $\sigma^2$ are the bandwidth and the white Gaussian noise power, respectively. Moreover, the achievable rate constraint is also important for the practical scenario, and it must satisfy the minimum achieved rate threshold.

\subsection{Energy consumption model of UAV}

\par The total UAV energy consumption can be divided into two parts. One is the energy consumption about communications, i.e., the modulation and demodulation of signals as well as the radiation \cite{8663615}. The other is the energy required for the UAV to overcome gravity and propulsion, which ensures that the UAV can hover and fly. It is worth noting that the communication energy consumption is much smaller than the energy consumed to overcome the gravity and propulsion \cite{8663615}. Therefore, according to \cite{8663615}, the energy consumption model in two-dimensional (2D) horizontal space of a rotary-wing UAV is proposed as follows:

\begin{equation}
\label{UAV-2D-Power}
\begin{split}
P(V)=&P_{B}\left(1+\frac{3{V}^2}{U_{tip}^{2}}\right)+P_{I}\left(\sqrt{1+\frac{{V}^4}{4v_{0}^{4}}}-\frac{{V}^2}{2v_{0}^{4}}\right)^{\frac{1}{2}}\\&+\frac{1}{2}d_{0}\rho sAV^{3}
\end{split}
\end{equation}

\noindent where $P_B$ and $P_I$ are the blade profile power and the induced power in hovering status, respectively, which are both constants. $U_{tip}$, $v_{0}$, $d_{0}$ and $\rho$ represent the tip speed of the rotor blade, the average rotor induction speed hovering, airframe drag ratio and air density, respectively. $s$ and $A$ represent the rotor solidity and the area of the rotor disk, respectively. Moreover, we ignore the effects from the acceleration or deceleration of UAV since they cost a tiny time \cite{8663615}.

\par Furthermore, this model is extended in 3D space, and the corresponding energy consumption model can be approximately expressed as follows \cite{8918497}:

\begin{equation}
\label{UAV-3D-Power}
\begin{split}
E(V)\approx &\int_{0}^{T} P(V(t))dt+\frac{M_{UAV}({V(T)}^2-{V(0)}^2)}{2}\\&+M_{UAV}g(Z(T)-Z(0))
\end{split}
\end{equation}
\noindent where $V(t)$ represents the instantaneous UAV speed at time $t$, and $T$ is the total flight time. $M_{UAV}$ and $g$ represent the mass and the gravitational factor, respectively. Moreover, $Z(T)$ and $Z(0)$ represent the end and start altitude, respectively.

\subsection{Obstacle model}
\par In this paper, we assume that the sizes and positions of the obstacles are known, since they can be measured by the 3D point cloud and remote sensing technique. Specifically, we consider several obstacles and the model of them are referred in \cite{9795684}, and the simplified model of the obstacles can be described as follows: 

\begin{equation}
\label{Obstacles}
\begin{split}
Ob(x^o, y^o)=&\sum_{m=1}^{G}h_m\\&\cdot \exp \left(-\left[\left(\frac{x^o-x^l_m}{x^s_m}\right)^2-\left(\frac{y^o-y^l_m}{y^s_m}\right)^2\right]\right)
\end{split}
\end{equation}
\noindent where $(x^o, y^o)$ is the projection of a certain point on a horizontal plane, and $Ob(x^o, y^o)$ is the corresponding obstacle altitude of this point. $(x^l_m, y^l_m)$ controls the horizontal center of the $m$th obstacle. $h_m$ is the altitude parameter of the $m$th obstacle. $x^s_m$ and $y^s_m$ jointly control the slope of the $m$th obstacle, and $G$ is the number of obstacles. Then, WDs will be deployed randomly on the ground area without obstacles.

\subsection{Multi-objective optimization problem}

\par
\begin{figure}[t]
	\centering{\includegraphics[width=2.5in]{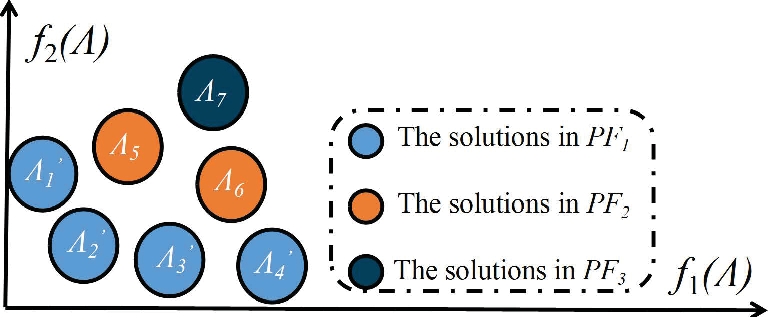}}
	\caption{Solution distribution of an MOP with two optimization objectives.}
	\label{figMOP}
\end{figure}

 In a multi-objective optimization problem (MOP), the solutions can be compared according to the Pareto dominance. This often occurs when comparing two solutions $\Lambda_1$ and $\Lambda_2$: $f_a(\Lambda_1)<f_a(\Lambda_2)$ but $f_b(\Lambda_1)>f_b(\Lambda_2)$. Thus, the concept of Pareto optimality is introduced to solve this problem. We define $\Lambda_3$ dominates $\Lambda_4$, recorded as $\Lambda_3\prec \Lambda_4$, if and only if $\Lambda_3$ is better than $\Lambda_4$ in at least one objective, which can be expressed mathematically as follows:
\begin{subequations}
	\label{eqMOP}
	\begin{align}
	f_o(\Lambda_3)\leq f_o(\Lambda_4), \forall o\in {1, 2, 3,...,obj}\\
	f_o(\Lambda_3)< f_o(\Lambda_4), \exists o\in {1, 2, 3,...,obj}		
	\end{align}
\end{subequations}
\noindent where $obj$ is the number of objective functions. If there is no $\Lambda$ satisfying $\Lambda \prec \Lambda^\prime$, $\Lambda^\prime$ can be called the non-dominated solution (also can be called the Pareto optimal solution). Generally, $\Lambda^\prime$ is not a single solution. In other words, the optimal solution for an MOP is a non-dominated solution set instead of a single solution. Then, the corresponding set in the objective space is the Pareto front (PF). More intuitively, Fig. \ref{figMOP} shows an example, which is a solution distribution of an MOP with two optimization objectives. As can be seen, there are several solutions which do not dominate each other, (e.g., ${\Lambda_1}^\prime$, ${\Lambda_2}^\prime$, ${\Lambda_3}^\prime$, ${\Lambda_4}^\prime$), and the algorithm for MOP will iteratively find several non-dominated solution sets and PFs. Then, the optimal PF, i.e., $PF_1$ in Fig. \ref{figMOP} is the non-dominated solutions obtained by the algorithm. The decision makers need to select a proper solution from $PF_1$ according to the realistic requirements.
\section{Joint power and 3D trajectory optimization of UAV}
\label{Problem statement}

\subsection{Problem formulation}

\par In this work, we aim to maximize the number of WDs to be served by UAV. Specifically, the UAV flies within the altitude limits. It is assumed that the required energy of each WD is the same and the collected data from each WD is of the same size. Moreover, the UAV needs to hover at a point for covering a set of WDs within its reasonable coverage range and then moves to next hovering point to do the same thing. It is worth noting that a UAV is able to cover multiple WDs at each hovering point simultaneously if these WDs are within its reasonable coverage range. However, there are four challenges for the coverage process of the UAV. \textbf{First}, in order to cover more WDs, hovering points should cover as different WDs as possible. However, such hovering points may be far apart, which may cause the increasing of flying energy consumption of UAV. \textbf{Second}, the UAV needs to hover to cover the corresponding WDs, which may cause extra time consumption. \textbf{Third}, the times of charging process and the data collection process will be different, which will reduce time efficiency. \textbf{Finally}, the obstacles should be considered in practical scenarios, and avoiding these obstacles will cause extra flight energy consumption of UAV.

\par Therefore, in the scenario with obstacles, the ultimate goal is to improve the energy utilization efficiency in a UAV-enabled WPCN. In other words, we aim to have the UAV to cover as many WDs as possible with the fixed hovering number by a most energy-saving way. Note that improving the energy utilization efficiency can be directly influenced by three optimization objectives as follows.

\par \textbf{\emph{Optimization objective 1: Maximize the total number of the covered WDs.}} To improve the energy utilization efficiency, more WDs should be covered by the UAV. Accordingly, the first objective function can be formulated as follows:
\begin{equation}
\label{f1}
\begin{split}
f_1(Q)= \sum_{j=1}^{M}Ac_j\\
\end{split}
\end{equation}
\noindent where $Ac_j$ is the number of the covered WDs at $q_j$. $Q$ represents a set of coordinates of all UAV hovering points which can be expressed as $[x_{1}, y_{1}, z_{1},..., x_{M}, y_{M}, z_{M}]$, where $(x, y, z)\in Q$ and $M$ is the total hovering number of a UAV.

\par \textbf{\emph{Optimization objective 2: Maximize the UAV time efficiency.}} The time efficiency can be influenced by three parts directly, which are the UAV total hovering time, the total time difference between the charging process and data collection process, and the penalty time. These three parts can synthesize an optimization objective to affect the time efficiency by a linear weighting method to make these three parts to be as small as possible, and the explicit analysis is as follows.
\par As mentioned above, during the hovering status, the UAV needs to collect data from the WDs one by one and charge all covered WDs simultaneously, which means the that charging process and data collection process are concurrent. Generally, the distance thresholds for data transmission and charging should be different. However, in this paper, our goal is to collect data from WDs and charge them simultaneously, which means that only considering collecting data from WDs or charging is meaningless. Thus, the distance threshold should be set as the smaller value of the distance thresholds for data collection and charging. Furthermore, the distance threshold of charging is smaller than that of data collection, since data and energy receivers operate with very different power sensitivity (e.g., $10$ dbm for energy receivers versus $60$ dbm for data receivers) \cite{liu2012wireless}. The above reason means that the transmission distance for charging is less than that for data collection, which can also be verified in several previous works. For data collection, according to \cite{DBLP:journals/tsp/ShenCGZZ20}, RF signals are capable of propagating a long distance, thus WIT can achieve long distance communication. For example, the authors in \cite{DBLP:journals/tsp/ShenCGZZ20} utilize the UAVs to achieve WIT over $100$ m. While for charging, according to \cite{DBLP:journals/cn/LiSWLLKL22}, the charging distance threshold is $30$ m. Thus, $d_{max}$ is set as the distance thresholds for data collection and charging. In other words, the maximum altitude of hovering points is set as $d_{max}$.

\par Assume that $e$ is the required energy of each WD and $u$ is the size of collected data from each WD for the UAV. Moreover, we assume that $e$ and $u$ are the same for each WD. Then, the higher transmission power may cause the lower time of the data collection process if the channel condition remains unchanged, which can influence the hovering time. For improving the time efficiency, we add the transmission power allocation of UAV at each hovering point into the solution space. Specifically, assume $Pc_j$ is the transmission power of the UAV at $q_j$, and $Pc=\{Pc_1,.., Pc_j,.., Pc_M\}$. Similarly, the transmission power of covered WDs is $Pt=\{Pt_1,.., Pt_i,.., Pt_{f_1}\}$, $f_1$ is the number of the covered WDs, which is the first optimization objective function. Then, the charging time between hovering point $q_j$ and $S_i$ is $Tc^i_j=e/Pr_{ij}$, and the data collection time is $Tdc^i_j=u/R_{ij}$. Furthermore, the charging time at $q_j$ is the maximum of those of all covered WDs, and the data collection time at $q_j$ is the sum of those of all covered WDs, which can be expressed as follows:
\begin{subequations}
	\label{Tc-Tdc}
	\begin{align}
	&Tc_j=\max\{Tc^1_j, ...Tc^i_j, ...Tc^{Ac_j}_j\}\\[-0.5mm]
	&\qquad\quad Tdc_j=\sum_{i=1}^{Ac_j}Tdc^i_j
	\end{align}
\end{subequations}
\noindent Moreover, the minimum achievable rate must satisfy the minimum achievable rate threshold, i.e. $R_{ij} \geq R_{th}$. In addition, since the charging process and the collecting data process are concurrent, the hovering time when UAV is at hovering point $q_j$ is maximum of the charging time and the data collection time, which can be formulated as follows:

\begin{equation}
\label{Th}
\begin{aligned}
Th_j = \max\{Tc_j, Tdc_j\}
\end{aligned}
\end{equation}

\par In addition, due to the above mentioned charging and data collection processes, there may be a time difference between the charging process and the data collection process, and such a time difference will cause unnecessary waste of energy. For example, if the time of charging is longer than the time of data collection, the transmission power of data collection can be given a smaller value to make the charging process and the data collection process finish at the same time, so that saving the energy consumption of data collection process. The time difference at $q_j$ between the charging and data collection process can be expressed as follows:
\begin{equation}
\label{Td}
\begin{aligned}
Td_j = |Tc_j-Tdc_j|
\end{aligned}
\end{equation}

\noindent Accordingly, the second objective function can be designed as follows:
\begin{equation}
\label{f2}
\begin{split}
f_2(Q, Pc, Pt)= \sum_{j=1}^{M} Th_j+\sum_{j=1}^{M} Td_j+n\cdot Pe\\
\end{split}
\end{equation}
\noindent where $n$ is the number of hovering points which cannot cover any WD, $Pe$ is a constant, and $n\cdot Pe$ is regarded as the penalty time. Note the smaller value of $f_2$ represents that the time efficiency is higher.
	\begin{remark}
		\label{nPe}
		{Consider an exceptional case that there is no charging and data collection processes. Then, the total hovering time and total time difference is equal to $0$, and there is no doubt that this value will be the optimal solution since we aim to achieve the minimum value of the UAV total hovering time and the total time difference. However, this will lead to the UAV to find the hovering points that cannot cover any WD, which does not make sense. Moreover, if we set a strong constraint that each hover point can cover one WD at least, which means that the algorithm may re-generate a solution when the abovementioned constraint is not satisfied, it will waste a lot of search time of the algorithm so that increasing the cost. Thus, we set a penalty time that the UAV will also hover at a hovering point for $Pe$ time if this hovering point cannot cover any WD.}
	\end{remark}

\color{black}

\par
\begin{figure}[t]
	\centering{\includegraphics[width=3.5in]{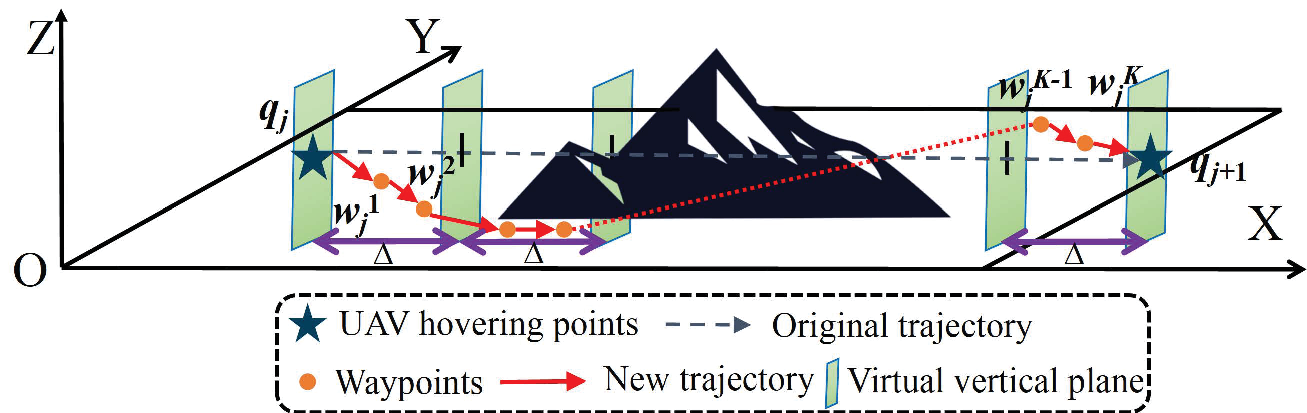}}
	\caption{An example of re-describing the 3D flying trajectory from $q_j$ to $q_{j+1}$.}
	\label{waypoints}
\end{figure}

 \textbf{\emph{Optimization objective 3: Minimize the total flying distance of UAV while avoiding obstacles.}} As mentioned above, the UAV needs to avoid obstacles, which means that the energy consumption of UAV may increase. According to the Eqs. (\ref{UAV-2D-Power}) and (\ref{UAV-3D-Power}), the flying energy consumption is related to the total flying distance due to the constant speed of UAV. In other words, to minimize the flying energy consumption can be transformed to minimize the flying distance. Moreover, since the considered scenario exists several obstacles, it is possible that the 3D flight trajectory from one hovering point to another one may collide them. Thus, we introduce the waypoints to avoid the obstacles which can be seen in Fig. \ref{waypoints}, and the explicit analysis of obstacle avoiding is presented in Section \ref{Regular}. As shown in the figure, we introduce $K$ waypoints for each of two adjacent hovering points to re-describe a 3D flight trajectory so that the UAV can bypass or fly over these obstacles. Moreover, $z(x, y)$ is the altitude of hovering point at $(x, y, 0)$, $z^*(x, y)$ is the altitude of waypoint at $(x, y, 0)$, and $Ob(x, y)$ is the altitude of obstacles at $(x, y, 0)$. Then, the hovering points and the waypoints must satisfy the following constraints to let UAV avoid obstacles: $Ob(x, y) < z(x, y)$, $Ob(x, y) < z^*(x, y)$. Assume that $w_j^k$ is the $k$th waypoint between the hovering point $q_j$ and $q_{j+1}$, and $Q^*$ is a set of 3D coordinates of all waypoints which can be expressed as $Q^*= [x_{1}^{*1}, y_{1}^{*1}, z_{1}^{*1},..., x_{1}^{*K}, y_{1}^{*K}, z_{1}^{*K},..., x_{M}^{*K}, y_{M}^{*K}, z_{M}^{*K}]$, where $(x^*, y^*, z^*)\in Q^*$. Then, the third objective function can be designed as follow:
\begin{equation}
\label{f3}
\begin{split}
f_3(Q, Q^*)= \sum_{j=1}^{M} (Ed_{j}^1+Ed_{j+1}^K)+\sum_{j=1}^{M}\sum_{k=1}^{K-1} Ed_j^{k,k+1}
\end{split}
\end{equation}
\noindent where $Ed_{j}^k=\sqrt{(x_j-x^*_k)^2+(y_j-y^*_k)^2+(z_j-z^*_k)^2}$ is the Euclidean distance between $q_j$ and $w_{j}^{k}$. $Ed_j^{k,k+1}=\sqrt{(x^*_{k+1}-x^*_k)^2+(y^*_{k+1}-y^*_k)^2+(z^*_{k+1}-z^*_k)^2}$ is the Euclidean distance between $w_{j}^{k}$ and $w_{j+1}^{k}$. Moreover, we set $q_{M+1}=q_1$ to ensure the UAV can return to the start point. Note that the maximum altitude of waypoints is allowed to be larger than $d_{max}$, since the waypoints are used to avoid obstacles instead of considering charging or data collection. Thus, we assume that the maximum altitude of waypoints is set as $Z_{max}$.
\noindent Then JUPTTOP can be formulated as follows: 
\begin{subequations}
	\label{JUPTTOP}
	\begin{align}
	\text{(JUPTTOP)}
	\mathop{\text{min}}\limits_{\{Q\},\{Q^*\}, \{Pc\}, \{Pt\}}~f&=\{-f_1, f_2, f_3\}\\
	\text{s.t.} 
	\qquad \mathcal{C}_1: &X_{min} \leq x \leq X_{max}\\
	\qquad \mathcal{C}_2: &Y_{min} \leq y \leq Y_{max}\\
	\qquad \mathcal{C}_3: &Z_{min} \leq z(x, y) \leq d_{max}\\
	\qquad \mathcal{C}_4: &Ob(x, y) < z(x, y) \\
	\qquad \mathcal{C}_5: &X_{min} \leq x^* \leq X_{max}\\
	\qquad \mathcal{C}_6: &Y_{min} \leq y^* \leq Y_{max}\\
	\qquad \mathcal{C}_7: &Z_{min} \leq z^*(x, y) \leq Z_{max}\\
	\qquad \mathcal{C}_8: & Ob(x, y) < z^*(x, y) \\
	\qquad \mathcal{C}_9: &Pc_{min} \leq Pc \leq Pc_{max}\\
	\qquad \mathcal{C}_{10}: &Pt_{min} \leq Pt \leq Pt_{max}\\
	\qquad \mathcal{C}_{11}: &R_{ij} \geq R_{th}
	\end{align}
\end{subequations}
\noindent where $f_1$, $f_2$, and $f_3$ of Eq. (\ref{JUPTTOP}a) denote the total number of the covered WDs, the time efficiency, and the total flying distance of UAV while avoiding obstacles, respectively, and these optimization objectives are trade-offs. For instance, if the altitudes of hovering points of the UAV are higher, it may face fewer obstacles so that reducing the flying distance (i.e., reducing the value of $f_3$). However, this will take extra hovering time of the UAV (i.e., increasing the value of $f_2$), since the distances between the hovering points and WDs are increasing. Thus, these optimization objectives should be jointly considered. Moreover, the optimization directions of the three optimization objectives are different. Specifically, $f_1$ is a maximization objective function while $f_2$ and $f_3$ are both minimization objective functions. Thus, Eq. (\ref{JUPTTOP}a) is formulated as a minimization problem. In addition, the solutions of optimization objective functions are related. Specifically, they have the same common solution part, i.e., the hovering points ${Q}$. For $f_1$, we use the hovering points to calculate the distances between the hovering points and WDs, then compare them with the maximum coverage threshold to judge whether the WDs are covered. For $f_2$, we also use the hovering points to calculate the distances between the hovering points and WDs, then calculate the charging time and the data collection time. Furthermore, we can calculate the hovering time and the time difference between the charging time and the data collection time, and further calculate the time efficiency. For $f_3$, the hovering points influence the total distance of UAV, since the hovering points are included in the UAV 3D trajectory.

\subsection{Problem conversion and analysis}

\par The 3D trajectory optimization problem is included in JUPTTOP, which is an NP-hard problem with strict constraints considering the obstacles, and the solution of the problem has a large solution search space. Moreover, the dimension of the problem is potential to be large which can increase the time cost of solving the problem. If the number of hovering points or the waypoints increases, the problem will be a large-scale optimization problem. For example, if the number of hovering points is set as $20$, the number of waypoints is set as $20$, and there are $300$ covered WDs, the dimension of the solution is $(4\times 20+3\times 20\times 20+300)=1580$, which is difficult to solve. In addition, the solution of the formulated JUPTTOP has different physical meanings and the dimension of the solution is not preset, which can further increase the difficulty. Thus, it is divided into two sub optimization problems that are UPAOP and UTTOP, respectively. Specifically, UPAOP is formulated as follows: 
\begin{subequations}
	\label{UPAOP}
	\begin{align}
	\text{(UPAOP)}\quad
	\mathop{\text{min}} \limits_{\{Q\}, \{Pc\}, \{Pt\}}&\quad f_{UPAOP}=\{-f_1, f_2\}\\
	\text{s.t.}\qquad &\quad \mathcal{C}_1-\mathcal{C}_4, \mathcal{C}_9-\mathcal{C}_{11}
	\end{align}
\end{subequations}
\noindent Similarly, UTTOP is formulated as follows:
\begin{subequations}
	\label{UTTOP}
	\begin{align}
	\text{(UTTOP)}\quad
	\mathop{\text{min}} \limits_{\{Q^*\}} \quad&f_{UTTOP}=f_{3}\\
	\text{s.t.} \quad& \mathcal{C}_5-\mathcal{C}_8
	\end{align}
\end{subequations}
\noindent By the conversions above, the original MOP is divided in an another MOP and a single objective problem. It can be seen from Eq. (\ref{UTTOP}a) that UTTOP is a multi-segment 3D trajectory optimization problem. For simplicity of analysis, we only analyze one segment 3D trajectory. It can be described as the problem of finding the optimal flight paths for the known start point and destination point in the flight area with obstacles, which has been proven as an NP-hard problem \cite{20213D}, since it has indefinite possible trajectories from the known start point to destination point. Therefore, UTTOP is NP-hard, and JUPTTOP is also NP-hard.

\section{Algorithm for UPAOP}
\label{Algorithm for UPAOP}

\par The formulated UPAOP is an MOP with constraints and there is a trade-off in UPAOP. Thus, a feasible strategy is to adopt multi-objective evolutionary algorithms to solve it, since the performances of multi-objective evolutionary algorithms are superior in solving the optimization problem with constraints. Moreover, they do not need derivative information in the solution process. However, the conventional non-dominated sorting genetic algorithm-II (NSGA-II) cannot satisfy the requirement of UPAOP, and the reasons are stated as follows: \textbf{First}, the random initialization in conventional NSGA-II fails to make good use of the prior information. \textbf{Second}, the solution dimension of the formulated UPAOP is not preset since the solution includes the $Pt$, and the dimension of $Pt$ is related to the total number of covered WDs, which is an optimization objective. Thus it motives us to propose an effective NSGA-II-KV for dealing with UPAOP.

\subsection{Conventional NSGA-II}

\par Among plenty of multi-objective evolutionary algorithms, NSGA-II is a multi-objective version of the genetic algorithm (GA). Thus, the main principle of conventional NSGA-II is similar to GA, which regards each solution of the optimization problem as a chromosome and the quality of a chromosome is evaluated by its fitness function value. Then, the crossover and mutation operations are used to update the solutions iteratively to find the best solution. Moreover, NSGA-II uses the Pareto optimal set to replace a single solution through calculating the Pareto dominance. The main steps of conventional NSGA-II are introduced in \cite{DBLP:journals/tec/Jensen03a}.

\subsection{Proposed NSGA-II-KV}

\par As analyzed above, UPAOP is an MOP whose solutions are the UAV hovering positions and the transmission powers of UAV and WDs. Moreover, UAV can hover any positions except the obstacles. Moreover, the dimension of transmission powers of WDs is not fixed. Thus, UPAOP is an MOP with a large solution search space and constraints. Conventional NSGA-II generates the initial chromosome randomly, and such method may set the UAV in the obstacles, hence making the algorithm fall into local optimum or making the UAV fall into the obstacles. Thus, the above reasons motive us to propose NSGA-II-KV to solve UPAOP better.

\subsubsection{K-means initialization operator}
\par In conventional NSGA-II \cite{DBLP:journals/tec/Jensen03a}, initial chromosomes generated randomly without prior knowledge are hard to inspire offspring chromosomes. However, a high quality heuristic function can improve the search efficiency \cite{9485039}, so that making each hovering point of UAV can cover at least one WD. Thus, $K$-means initialization operator is embedded into initialization process to improve the search efficiency. Specifically, regard the coordinates of WDs as the sample points, then generate $M$ cluster center. The initial 2D horizontal coordinates of UAV is the 2D horizontal coordinates of cluster centers. The main steps of the $K$-means initialization operator are shown in Algorithm \ref{k-means}.
\begin{algorithm}[t]
	\caption{$K$-means initialization operator} 
	\label{k-means}
	Obtain the 2D horizontal coordinates of WDs;\\
	Generate $M$ 2D horizontal coordinates cluster centers according to \cite{DBLP:journals/access/StiawanSSIAAB21};\\
	Initialize the coordinates of these cluster centers as the UAV 2D horizontal coordinates;\\
	Initialize the UAV initial altitudes of each hovering points randomly, which are subjected to the constraints.	
\end{algorithm}

\subsubsection{Variable dimension mechanism}
\par
\begin{figure}[htb]
	\centering{\includegraphics[width=3.5in]{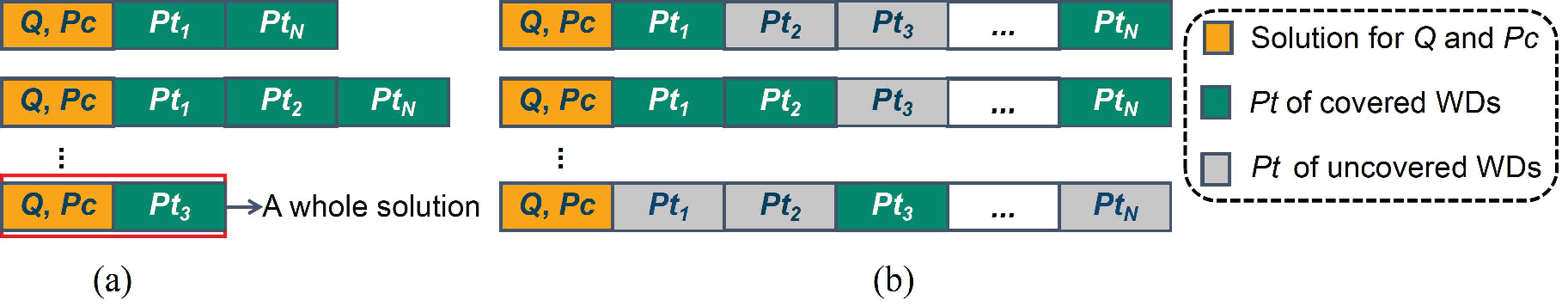}}
	\caption{Principle of variable dimension mechanism. (a) Original population. (b) Modified population after variable dimension mechanism.}
	\label{Variable dimension}
\end{figure}

 In UPAOP, $Pt$ is the part of a solution, and the dimension of $Pt$ can be influenced by the total number of covered WDs, which is an optimization objective and can be changed iteratively. Thus, we introduce a variable dimension mechanism to deal with the change of the dimension. Specifically, Fig. \ref{Variable dimension} illustrates the main principle of the variable dimension mechanism. Fig. \ref{Variable dimension}(a) is the original population and Fig. \ref{Variable dimension}(b) is the modified population after variable dimension mechanism. Specifically, this mechanism adds $Pt$ of uncovered WDs into the solution, so as to ensure the algorithm crossover and mutation. However, the $Pt$ of uncovered WDs cannot participate in the calculation of objective functions. 

\par Assume that $G_{max}$ is the maximum number of iterations. Then the overall algorithm processes of NSGA-II-KV are shown in Algorithm \ref{NSGA-II-KV}.
\begin{algorithm}
	\caption{NSGA-II-KV} 
	\label{NSGA-II-KV}
	\textbf{Define the fitness function:} $f_{UPAOP}(Q, Pc, Pt)$;\\
	Initialize the population according to Algorithm \ref{k-means};\\
	Initialize non-dominated sorting and crowding sorting of population;\\
	\While{$it<G_{max}$}
	{
		Crossover and mutation to produce offspring UAV hovering positions;\\
		Combine parent and offspring population;\\
		Compute non-dominated sorting and crowding distance of offspring population;\\
		Elite retention to generate next generation population;\\
		$it=it+1$;
	}
\end{algorithm}

\subsection{Complexity and convergence analysis of NSGA-II-KV}

\subsubsection{Analysis of complexity} The improvement of the NSGA-II-KV is to embed a conventional $K$-means algorithm into the initialization of the conventional NSGA-II, and introduce a variable dimension mechanism to ensure the algorithm to deal with the dimension length change condition. First, the variable dimension mechanism is applied for the NSGA-II-KV and other comparison algorithms, thus the computation complexity of the variable dimension mechanism can be ignored. Second, since there is no iteration in this initialization, the complexity of NSGA-II-KV is the sum of the complexity of the NSGA-II and $K$-means algorithm. Assume that the population size of the NSGA-II-KV is $NP$, and the maximum iteration of $K$-means algorithm is also set as $G_{max}$. Then, the computation complexity of the $K$-means algorithm can be computed as $\mathcal{O}(N\cdot M\cdot NP\cdot G_{max}\cdot I_1)$ \cite{DBLP:journals/access/StiawanSSIAAB21}, where $I_1$ is the processing time for calculating the distance between two hovering points. Moreover, the computation complexity of NSGA-II is $\mathcal{O}(NP^2\cdot G_{max}\cdot I_2)$ according to \cite{DBLP:journals/tec/Jensen03a},  where $I_2$ is the processing time for calculating the two objectives. In this paper, $NP>M$. Moreover, $I_2>N\cdot I_1$. The reason is that calculating the two objectives must calculate the distances between one hovering point and all WD positions. When UAV hovers at a position, it must calculate which WDs can be covered, i.e., the distance calculating will be repeated for $N$ times at least. Thus, $\mathcal{O}(NP^2\cdot G_{max}\cdot I_2)>\mathcal{O}(N\cdot M\cdot NP\cdot G_{max}\cdot I_1)$, and the overall computation complexity of NSGA-II-KV is $\mathcal{O}(NP^2\cdot G_{max}\cdot I_2)$.
\subsubsection{Analysis of convergence} The proposed NSGA-II-KV is an improved evolutionary algorithm based on the conventional NSGA-II, and the evolutionary algorithms will be sure to be convergent for any tolerance whatever for unimodal or multimodal functions, which is analyzed in \cite{zhao2007evolutionary} and the authors give two lemmas on the probability of locating in the promising area and the different distance from the initial solution to the optimal solution. Thus, the proposed NSGA-II-KV is convergent.
\label{Analysis of convergence}

\section{Algorithm for UTTOP}
\label{Algorithm for UTTOP}

\par The formulated UTTOP is still an NP-hard problem with the constraints, which means no algorithm can find the optimal solution in polynomial time. Thus, an alternative strategy is to adopt evolutionary algorithms for finding a feasible solution in limited time. Moreover, evolutionary algorithms have been proven that they have many advantages in solving NP-hard problems with the constraints. However, the conventional PSO cannot satisfy the solving requirements of UTTOP, since it is easy to fall into local dilemma due to the constraints. Therefore, a pretreatment method is proposed to make the algorithm satisfy the constraints, then we propose a PSO-NGDP to cope with the formulated UTTOP.

\subsection{Conventional PSO}
\par PSO is a popular meta-heuristic optimization method since it converges fast \cite{5675669}. Similar to general single objective evolutionary algorithms, it starts with the random initialization and iteratively finds the global optimal solution. In conventional PSO, the positions of a particle are regarded as a solution and it uses velocity updating to describe the change of the position. Moreover, in each iteration, the algorithm can memorize the individual best $P_{best}$ and the global best position $G_{best}$. The main update process of the population can be expressed as follows \cite{5675669}:

\begin{subequations}
	\label{PSO-equation}
	\begin{align}
	&v^b= \omega * v^b + C_{1} * r_{1} * (P_{best}^b - \chi^b)  + C_{2} * r_{2} * (G_{best} -\chi^b)\\
	&\qquad\qquad\qquad\qquad\qquad\chi^b = \chi^b + v^b
	\end{align}
\end{subequations}
\noindent where $v^b$ is the velocity of $b$ particle and $\chi^b$ is the position of $b$ particle. $C_1$ and $C_2$ are two acceleration coefficients reflecting the ability of learning, respectively. $r_{1}$, $r_{2}$ are two random numbers distributed in the range from $\left( 0, 1 \right)$, respectively. The main steps of conventional PSO can be found in \cite{5675669}.

\subsection{Pretreatment method}
\label{Regular}

\par
\begin{figure}[t]
	\centering{\includegraphics[width=3.5in]{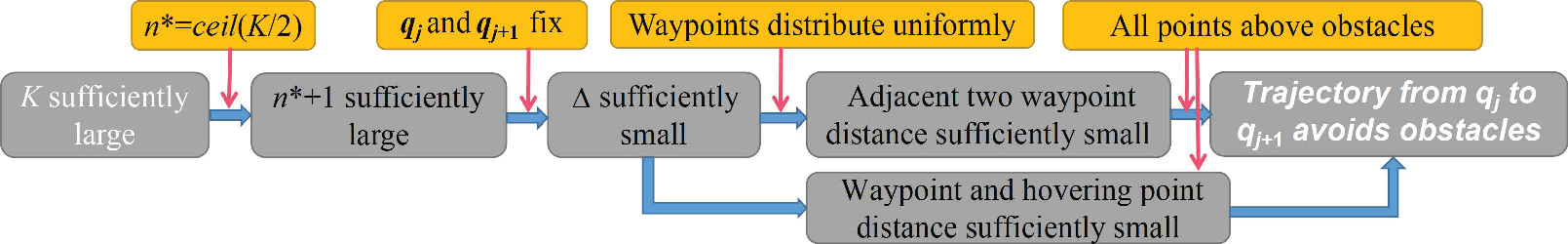}}
	\caption{Main processes of the pretreatment method.}
	\label{Dwaypoints}
\end{figure}

 To ensure that the UAV does not collide with the obstacles, we introduce the waypoints to bypass or fly over them. However, UAV may choose the any way around to achieve the goal, and it possibly chooses the farthest way. Thus, we consider a new method to ensure UAV flying to the destination with the less energy. For example, $q_j$ and $q_{j+1}$ are regarded as the start point and the destination point. Assume $K$ is the number of waypoints between $q_j$ and $q_{j+1}$, and the main processes of the pretreatment method are shown in Fig. \ref{Dwaypoints}. Then, the main steps of this method can be expressed in Algorithm \ref{Regular method}, which is detaled as follows.
\begin{algorithm}[htb]
	\caption{Pretreatment method}	
	\label{Regular method}
	Connect $q_j$ and $q_{j+1}$, which is recorded as $\sim(q_j, q_{j+1})$;\\
	Take $\sim(q_j, q_{j+1})$ as the normal and assume that $n^*=\lceil K/2 \rceil$, then make $n^*-1$ virtual vertical planes, and these $n^*-1$ planes can divide $\sim(q_j, q_{j+1})$ into $n^*$ segments with equal length $\Delta$. These virtual vertical planes are recorded as $\Omega_1, \Omega_2, ..., \Omega_{n^*-1}$;\\	
	Take $\sim(q_j, q_{j+1})$ as the normal, and make two virtual vertical planes through $q_j$ and $q_{j+1}$, which are recorded as $\Omega_0$ and $\Omega_{n^*}$, respectively. Then, these $n^*+1$ virtual vertical planes will generate ${n^*}$ subspaces;\\
	Divide $K$ waypoints into these ${n^*}$ subspaces uniformly. Judge whether the waypoints in each subspace are between the corresponding two planes. If not, we should project the position of the waypoint to the virtual vertical plane that is closest to the waypoint;\\
\end{algorithm}
\par Assume that all the waypoints are above the obstacles and $K$ is sufficiently large. Then, $n^*=\lceil K/2 \rceil$ is correspondingly large, which means that the virtual vertical plane number $n^*+1$ in Algorithm \ref{Regular method} is sufficiently large. The hovering points must be determined by solving UPAOP, which means that the distance between $q_j$ and $q_{j+1}$ is fixed as a constant, then there are two cases to be discussed. First, two adjacent waypoints are both in the same subspace. Second, two adjacent waypoints are in the adjacent subspaces, respectively. In any case of the two cases, since the distance between $q_j$ and $q_{j+1}$ is fixed, and the number of the virtual vertical planes is sufficiently large, then the distance between any two adjacent virtual vertical planes is sufficiently small. Moreover, the waypoints are divided into these subspaces uniformly. Thus, the distance between two adjacent waypoints can be sufficiently small. Then, the line that connecting the two adjacent waypoints can be regarded as obstacle avoidance. Similarly, the line that connecting the corresponding waypoint and hovering point can also be regarded as obstacle avoidance. In addition, since the trajectory from $q_j$ and $q_{j+1}$ is a sequential connection via $q_j$, waypoints and $q_{j+1}$, the trajectory from $q_j$ to $q_{j+1}$ can avoid the obstacles. Accordingly, the whole trajectory of the UAV can avoid the obstacles.

\subsection{Proposed PSO-NGDP}

\par The formulated UTTOP is a continuous problem with constraints. Thus, we propose a PSO-NGDP with four improved factors to solve the formulated UTTOP. PSO-NGDP is extended by conventional PSO, and the details are presented as follows.

\subsubsection{Normal distribution initialization}
\par Conventional PSO is initialized by a random distribution solution, such that leading to certain blindness of search directions. Moreover, PSO may converge slowly for solving large dimension problem. Thus, a high quality initialization can improve the convergence speed of the algorithm \cite{8527444}. Thus, we adopt the normal distribution instead of the random distribution in the initialization process, and the 3D coordinates $Q_j^{*k}=(x_j^{*k}, y_j^{*k}, z_j^{*k})$ of $w_j^k$ can be initialized as Eq. (\ref{PSONGDP-initial}), where $randn$ is a random number generated using Gaussian distribution $\mathcal CN(0, X_{max}/100)$. $C_3$ is a constant, $ind=3(j-1)$, and $\mod(ind+1, 3K)$ is to calculate the remainder.

\begin{figure*}[!t]
	\normalsize
	\setcounter{MaxMatrixCols}{20}
\begin{subequations}
	\label{PSONGDP-initial}
\begin{align}
x_j^{*k} = \frac{ randn\cdot X_{max}}{C_3} + \frac{3K-\mod(ind+1,3K)}{3K} x_j + \frac{\mod(ind+1,3K)}{3K} x_{j+1}\\
y_j^{*k} = \frac{ randn\cdot Y_{max}}{C_3} + \frac{3K-\mod(ind+2,3K)}{3K} y_j + \frac{\mod(ind+2,3K)}{3K} y_{j+1}\\
z_j^{*k} = \frac{ randn\cdot Z_{max}}{C_3} + \frac{3K-\mod(ind+3,3K)}{3K} z_j + \frac{\mod(ind+3,3K)}{3K} z_{j+1}
\end{align}
\end{subequations}

	\hrulefill
	\vspace*{4pt}
\end{figure*}

\subsubsection {Genetic mechanism}
\par Facing a large dimension problem UTTOP, the convergence rate of conventional PSO is limited, which motives us to combine PSO and GA to jump out of the local optimal solution. Specially speaking, we embed the crossover operation of GA into PSO to jump out of the local dilemma. The details can be expressed as follows. 
\par
\begin{figure}[t]
	\centering{\includegraphics[width=3.5in]{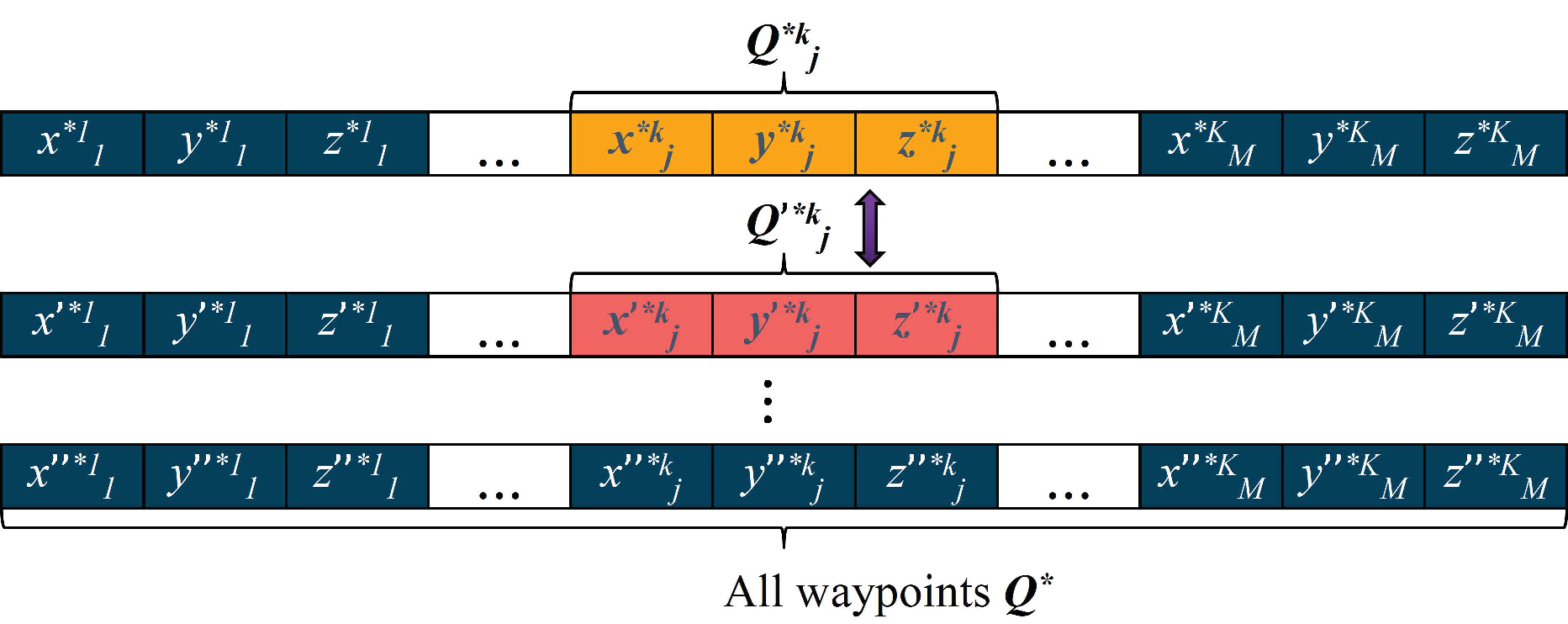}}
	\caption{Crossover process.}
	\label{Crossover}
\end{figure}

 \textbf{Crossover}: The main idea of the crossover operator is to exchange the 3D coordinates for two waypoints with the same index among the two particles, and the best waypoint may be set into the same particle. As shown in Fig. \ref{Crossover}, we change the corresponding waypoint to update the particle with a certain probability. The method can be expressed mathematically as follows:
\begin{equation}
\label{Crossover_a}
\begin{aligned}
& Q_j^{*k} \Leftrightarrow \begin{cases} Q_{j}^{\prime *k}, \quad r_{3} \leq r_{cro}\\
Q_j^{*k}, \quad \text{otherwise}
\end{cases}  
\end{aligned}
\end{equation}
\noindent where $Q^{\prime*}$ is the position of another particle. $r_3$ is the random number generated from $(0, 1)$, and $r_{cro}$ is the crossover probability. 

\subsubsection {Differential mechanism}
\par The PSO can converge fast due to its fewer parameters. However, it can lead to the algorithm to drop into local dilemma. Instead, other evolutionary algorithms such as differential evolution (DE) can jump out of the local optimal solution due to mechanisms such as mutation. Therefore, we combine the PSO with the mutation and natural selection mechanisms in the DE to make the overall algorithm jump out of the local optimum. The details of the process can be expressed as follows.

\par \textbf{Mutation:} In order to enhance the diversity of the population, we combine PSO with the mutation operation in DE. Such mechanism can generate the new solutions to make the algorithm jump out of the local optimal. It can be seen from Eq. (\ref{mutation}) that the mutation mechanism is embedded into PSO to update the positions. The method of generating new solution is as follows:
\begin{equation}
\label{mutation}
\begin{aligned}
&Q _j^{\prime *k}=Q_j^{*k}+F_{0} \cdot (Q_{j^\prime}^{*k} - Q_{j^{\prime\prime}}^{*k})
\end{aligned}
\end{equation}
\noindent where $j^\prime, j^{\prime\prime} \in\{j\neq j^\prime, j\neq j^{\prime\prime}, j^\prime\neq j^{\prime\prime}|1, 2, 3,..., M\}$. $F_0$ is the coefficient of variation operator.

\par \textbf{Natural selection:} The mutation operation can generate new particles. In order to maintain the population size, we introduce a natural selection mechanism. Specifically speaking, this mechanism adopts a greedy criterion to select the better particles and it can be expressed as follows:
\begin{equation}
\label{Natural selection}
\begin{aligned}
& Q_j^{*k}=\begin{cases} Q_j^{\prime *k} , \quad & f_{UTTOP}(Q_j^{\prime *k}) \le f_{UTTOP}(Q_j^{*k}) \\
Q_j^{*k}, \quad &\text{otherwise}
\end{cases}
\end{aligned}
\end{equation}
\noindent Moreover, the velocities in PSO can affect the solution updating iteratively, thus it is reasonable to update the corresponding velocities synchronously while executing Eqs. (\ref{Crossover_a}), (\ref{mutation}), and (\ref{Natural selection}). 

\subsubsection{Pursuit operator}
\par In the iterative process, we usually increase the altitude of the flight or recalculate the flight to avoid obstacles. Specifically, we consider to embed a pursuit operator iteratively into the method mentioned in Section \ref{Regular} to reduce the flight distance as much as possible. The pursuit operator utilizes both the global optimal position and the individual optimal position to update the solution. The update method of the solution can be expressed as follows:
\begin{equation}
\begin{aligned}
Q_j^{*k} = Q_j^{*k} + r_{4} \cdot \min ( | {G_{best}}_j^k- {P_{best}}_j^k|, r_{5} \cdot v_{max} ) 
\end{aligned}
\label{pursuit operator}
\end{equation}
\noindent where $v_{max}$ is the upper bound of the velocity of each particle. We use $|{G_{best}}_j^k-{P_{best}}_j^k|$ to express the disparity between the individual optimal waypoint and the global optimal waypoint. $r_{4}, r_{5}\in (0, 1)$ are two random numbers. Moreover, the upper and lower bound of the velocity can be dynamic tuning iteratively, which can be expressed as follows:
\begin{subequations}
	\begin{align}
	&v_{max} = v_1 -(v_1-v_2)*it/G_{max}\\
	&\qquad\quad v_{min} = -v_{max}
	\end{align}
	\label{dynamic tuning}
\end{subequations}
\noindent where $it$ is the current iteration, and $G_{max}$ is the maximum iteration. $v_1$ and $v_2$ are two constant. Therefore, the algorithm for UAV to deal with obstacles can be expressed as Algorithm \ref{deal_with_obstacles}.

\begin{algorithm}[t]
	\caption{Algorithm for UAV to deal with obstacles}
	\label{deal_with_obstacles}
	\For{j=1 to M}
	{
		\For{k=1 to K}{
			Set $AO= 0$ to count the number of failed obstacle avoidance relocations;\\
			\While{$Ob(x_j^{*k}, y_j^{*k})\geq z_j^{*k}$}
			{
				Update the coordinates of the waypoints according to Eq. (\ref{pursuit operator});\\
				$AO = AO + 1$;\\
				\If {$\mod(AO, 10)=0$}
				{$z_j^{*k}=z_j^{*k}+C_4*(AO/10)$;}
			}
		}
	}
\end{algorithm}
\par Assume the population size is $NP$,  and $C_4$ is a constant number, then the main steps of PSO-NGDP are shown in Algorithm \ref{PSONGDP}.

\subsection{Complexity and convergence analysis of PSO-NGDP}
\subsubsection{Analysis of complexity} For the proposed PSO-NGDP, the increasing of the computation complexity is mainly from the pretreatment method and the algorithm for UAV to deal with obstacles, which are shown as Algorithms \ref{Regular method} and \ref{deal_with_obstacles}, respectively. It can be seen that the computation complexity of Algorithms \ref{Regular method} is $\mathcal{O}(K)$. Moreover, there is a while loop in Algorithm \ref{deal_with_obstacles}, and we need to consider the worst situation. Since the maximum altitude of UAV is set as $Z_{max}$, and we assume the initial altitude of UAV is set as $Z_{min}$. Moreover, we assume that the altitude of obstacle is close to $Z_{max}$, which means Algorithm \ref{deal_with_obstacles} needs to execute $(Z_{max} - Z_{min})/C_4$ to jump out of the while loop at most. Then, the computation complexity for one iteration of Algorithms \ref{PSONGDP} is $\mathcal{O}((Z_{max}- Z_{min})\cdot M\cdot K/C_4)$ which is larger than $\mathcal{O}(K)$, thus the computation complexity of PSO-NGDP is $\mathcal{O}((Z_{max}- Z_{min})\cdot M\cdot K\cdot G_{max}/C_4)$.
\begin{algorithm}[t]
	\caption{PSO-NGDP}
	\label{PSONGDP}
	\textbf{Define the fitness function:}
	$f_{UTTOP}(Q^*);$\\
	Set the learning factors, the inertia weight and the maximum iteration number;\\
	Initialize the coordinates of the waypoints according to Eq. (\ref{PSONGDP-initial}) for each particle;\\
	Update the coordinates of the waypoints according to Algorithm \ref{Regular method} for each particle;\\
	Initialize the bound of the velocity according to Eqs. (\ref{dynamic tuning}a) and (\ref{dynamic tuning}b);\\
	Update the individual optimal solution $P_{best}$ and the global optimal solution $G_{best}$;\\
	\For{it=1 to $G_{max}$}
	{	
		\For{Each particle}{
			Crossover and mutation operator by using Eqs. (\ref{Crossover_a}) and (\ref{mutation}) for each waypoint;\\
			Natural selection according to Eq. (\ref{Natural selection});\\
			Update the coordinates of the waypoints according to Eq. (\ref{PSO-equation});\\
			Update the coordinates of the waypoints according to Algorithm \ref{Regular method};\\
			Execute Algorithm \ref{deal_with_obstacles} to deal with collision with obstacles;\\
			Calculate the fitness function and update ${P_{best}}$;\\
		}
	}
	Output $G_{best}$.\\
\end{algorithm}
\subsubsection{Analysis of convergence} PSO-NGDP is a variant of PSO, thus it is also convergent, and the reasons are similar to NSGA-II-KV. 
%
%

\section{Simulation results}
\label{Simulation results}

\par In this section, simulations are conducted based on Matlab to evaluate the performance of the proposed NSGA-II-KV and PSO-NGDP. In our considered scenarios, the space is set as 500 m $\times$ 500 m $\times$ 30 m. $Z_{max}$ and $Z_{min}$ are set as $30$ m and $5$ m, respectively. For the charging parameters, $Pc_{max}$ and $Pc_{min}$ are $3$ W and $1$ W, respectively. $\gamma$ and $\tau$ are set as $36$ and $30$, respectively \cite{liang2021charging}, and $d_{max}$ is set as $20$ m. We assume required energy of a WD $e$ is $10$ J. For the data collection parameters, $C=10$ and $D=0.6$. $\sigma^2=-100$ dBm and $\beta_{0}=-60$ dB. The size of collected data $u$ is $100$ Mbits and $R_{th}$ is set as $0.5$ Mbps. $Pt_{max}$ and $Pt_{min}$ are $1$ W and $0.1$ W, respectively. The bandwidth $B$ is set as $1$ MHz and $\kappa=0.2$. Note that the transmission power for data collection is much smaller than charging. Moreover, we consider 500 WDs for small size network and 1000 WDs for small size network, respectively. Furthermore, we consider three scenarios for small size networks (Scenario 1, Scenario 2 and Scenario 3) and large size networks (Scenario 4, Scenario 5 and Scenario 6), respectively, where the shapes and locations of obstacles are different\footnote{The parameters for obstacles are available at https://github.com/531669432/Parameter-Obstacles/tree/main.}. In addition, the numbers of hovering points are set as $10$ and $20$ for small and large size networks, respectively. Moreover, the population size and the maximum number of iterations of the two proposed algorithms are $50$ and $300$, respectively. 

\subsection{Results of UPAOP}
\par
\begin{figure}[htbp]
	\centering{\includegraphics[width=3.5in]{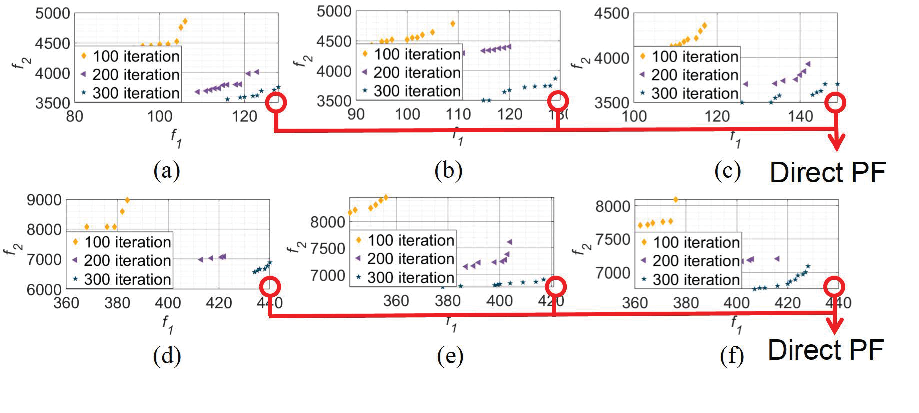}}
	\caption{Solution distributions of different iteration obtained by NSGA-II-KV. (a) Scenario 1. (b) Scenario 2. (c) Scenario 3. (d) Scenario 4. (e) Scenario 5. (f) Scenario 6.}
	\label{ParetoConvergence}
\end{figure}

\begin{figure}[t]
	\centering{\includegraphics[width=3.5in]{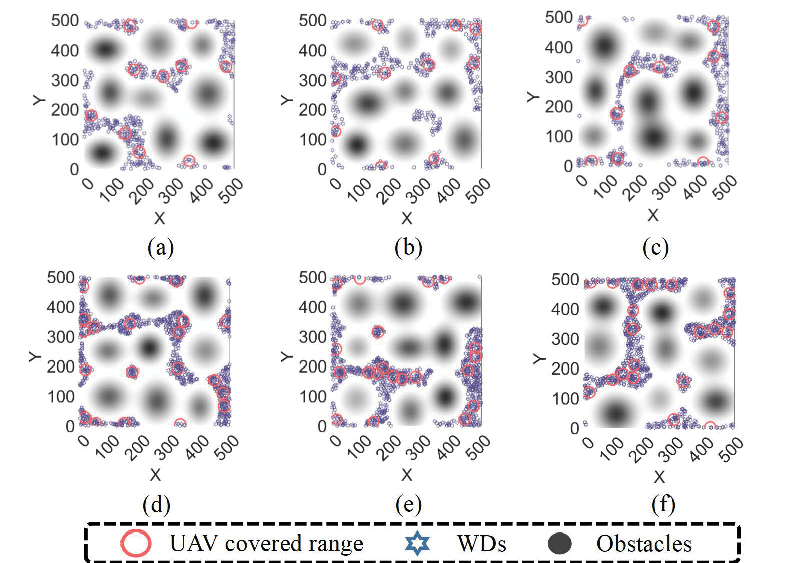}}
	\caption{Optimization results of UAV hoving points obtained by NSGA-II-KV for solving UPAOP. (a) Scenario 1. (b) Scenario 2. (c) Scenario 3. (d) Scenario 4. (e) Scenario 5. (f) Scenario 6.}
	\label{coverage}
\end{figure}

 Figs. \ref{ParetoConvergence}(a)-\ref{ParetoConvergence}(f) verify the convergence of NSGA-II-KV for Scenarios 1-6, respectively, and Figs. \ref{coverage}(a)-\ref{coverage}(f) demonstrate the vertical view of UAV scheduling and the covered WDs obtained by NSGA-II-KV, while the grey parts in figures are the obstacles, and the darker parts represent the higher obstacles. It can be seen from Figs. \ref{coverage}(a)-\ref{coverage}(f) that each hover point can cover at least one WD, i.e., $n$ is equal to $0$. Thus, $n\cdot Pe$ is equal to $0$, which cannot influence the value of $f_2$.
\par Then, we introduce the $K$-means algorithm \cite{DBLP:journals/access/StiawanSSIAAB21}, strength Pareto evolutionary algorithm-II (SPEA-II)\cite{DBLP:journals/tec/Jensen03a}, multi-objective dragonfly algorithm (MODA) \cite{2016Dragonfly}, multi-objective salp swarm algorithm (MSSA) \cite{2017Salp}, and conventional NSGA-II \cite{DBLP:journals/tec/Jensen03a} as comparison algorithms to solve UPAOP. Moreover, the uniform scheduling (US) and the random scheduling (RS) are also considered in the tests. Specifically, US means that the UAV is deployed uniformly in the 2D horizontal area and RS means the UAV is deployed randomly in 3D space. In US and $K$-means tests, the UAV flight height will be fixed at $5$ m to cover more WDs, which is the minimum altitude of UAV. Moreover, $Pt$ and $Pc$ are set as the maximum value in US, RS and $K$-means algorithm. 
\par
\begin{figure}[t]
	\centering{\includegraphics[width=3.5in]{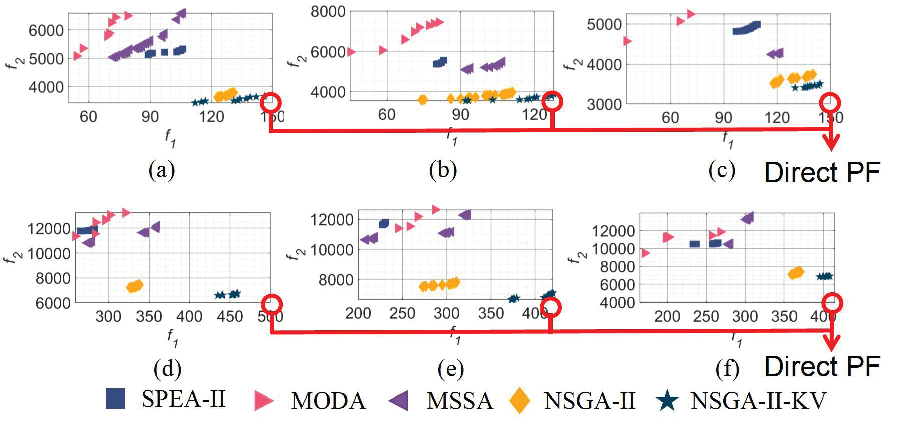}}
	\caption{Pareto optimization results obtained by NSGA-II-KV and other comparison algorithms for solving UPAOP. (a) Scenario 1. (b) Scenario 2. (c) Scenario 3. (d) Scenario 4. (e) Scenario 5. (f) Scenario 6.}
	\label{Pareto}
\end{figure}

 It can be seen from Figs. \ref{Pareto}(a)-\ref{Pareto}(f), the PF obtained by NSGA-II-KV are much closer to the direct PF than other comparison algorithms, which means that the proposed NSGA-II-KV performs better than other comparison algorithms for solving the proposed UPAOP. It can be seen from Table \ref{Select}, the proposed NSGA-II-KV obtains the best performances in small size networks. Among large size networks, the proposed NSGA-II-KV obtains the best performance in Scenario 5, and obtains the best performance of $f_1$ in Scenarios 4 and 6, while obtains the second best performance of the $f_2$. $K$-means algorithm obtains the best performances of the second objective function in Scenarios 4 and 6, since the altitudes of the hovering points are set as the minimum, which can reduce the hovering time, and can further reduce $f_2$. However, such setting will damage $f_3$ when UAV facing obstacles, since UAV needs to fly higher to avoid the obstacles which can increase the flying distance.
\begin{table*}[t]
	\scriptsize
	\centering
	\begin{center}
		\caption{Numerical statistical results obtained by NSGA-II-KV and other comparison algorithms for solving UPAOP}
		\label{Select}	
		\begin{tabular}{llllllllllllll}\hline
			&\multicolumn{1}{|c}{\textbf{Algorithm}}&\multicolumn{2}{|c|}{\textbf {Scenario 1}}&\multicolumn{2}{c|}{\textbf {Scenario 2}}&\multicolumn{2}{c|}{\textbf {Scenario 3}}&\multicolumn{2}{c|}{\textbf {Scenario 4}}&\multicolumn{2}{c|}{\textbf {Scenario 5}}&\multicolumn{2}{c}{\textbf {Scenario 6}}\\\hline

			\multirow{8}{*}{$f_1$}
			&\multicolumn{1}{|c}{\textbf{US}} &\multicolumn{2}{|c|}{5.00} &\multicolumn{2}{c|}{4.00} &\multicolumn{2}{c|}{3.00} &\multicolumn{2}{c|}{63.00} &\multicolumn{2}{c|}{47.00} &\multicolumn{2}{c}{92.00}\\

			&\multicolumn{1}{|c}{\textbf{RS}} &\multicolumn{2}{|c|}{25.00} &\multicolumn{2}{c|}{15.00} &\multicolumn{2}{c|}{18.00} &\multicolumn{2}{c|}{64.00} &\multicolumn{2}{c|}{55.00} &\multicolumn{2}{c}{46.00}\\

			&\multicolumn{1}{|c}{\textbf{$K$-means}} &\multicolumn{2}{|c|}{87.00} &\multicolumn{2}{c|}{83.00} &\multicolumn{2}{c|}{96.00} &\multicolumn{2}{c|}{374.00} &\multicolumn{2}{c|}{369.00} &\multicolumn{2}{c}{348.00}\\

			&\multicolumn{1}{|c}{\textbf{SPEA-II}} &\multicolumn{2}{|c|}{103.00} &\multicolumn{2}{c|}{83.00} &\multicolumn{2}{c|}{106.00} &\multicolumn{2}{c|}{278.00} &\multicolumn{2}{c|}{230.00} &\multicolumn{2}{c}{262.00}\\
			
			&\multicolumn{1}{|c}{\textbf{MODA}} &\multicolumn{2}{|c|}{73.00} &\multicolumn{2}{c|}{73.00} &\multicolumn{2}{c|}{71.00} &\multicolumn{2}{c|}{283.00} &\multicolumn{2}{c|}{258.00} &\multicolumn{2}{c}{257.00}\\

			&\multicolumn{1}{|c}{\textbf{MSSA}} &\multicolumn{2}{|c|}{89.00} &\multicolumn{2}{c|}{106.00} &\multicolumn{2}{c|}{121.00} &\multicolumn{2}{c|}{276.00} &\multicolumn{2}{c|}{305.00} &\multicolumn{2}{c}{306.00}\\

			&\multicolumn{1}{|c}{\textbf{NSGA-II}} &\multicolumn{2}{|c|}{129.00} &\multicolumn{2}{c|}{107.00} &\multicolumn{2}{c|}{122.00} &\multicolumn{2}{c|}{337.00} &\multicolumn{2}{c|}{306.00} &\multicolumn{2}{c}{366.00}\\
			
			&\multicolumn{1}{|c}{\textbf{NSGA-II-KV}} &\multicolumn{2}{|c|}{\textbf{134.00}} &\multicolumn{2}{c|}{\textbf{114.00}} &\multicolumn{2}{c|}{\textbf{144.00}} &\multicolumn{2}{c|}{\textbf{435.00}} &\multicolumn{2}{c|}{\textbf{376.00}} &\multicolumn{2}{c}{\textbf{395.00}}\\\hline

			\multirow{8}{*}{$f_2$}
			
			&\multicolumn{1}{|c}{\textbf{US}} &\multicolumn{2}{|c|}{8062.63} &\multicolumn{2}{c|}{8122.11} &\multicolumn{2}{c|}{9039.07} &\multicolumn{2}{c|}{14558.11} &\multicolumn{2}{c|}{15002.27} &\multicolumn{2}{c}{14714.09}\\

			&\multicolumn{1}{|c}{\textbf{RS}} &\multicolumn{2}{|c|}{9386.24} &\multicolumn{2}{c|}{6167.49} &\multicolumn{2}{c|}{6194.56} &\multicolumn{2}{c|}{14610.80} &\multicolumn{2}{c|}{14654.83} &\multicolumn{2}{c}{14931.28}\\

			&\multicolumn{1}{|c}{\textbf{$K$-means}} &\multicolumn{2}{|c|}{5021.57} &\multicolumn{2}{c|}{3738.53} &\multicolumn{2}{c|}{3865.92} &\multicolumn{2}{c|}{\textbf{6505.15}} &\multicolumn{2}{c|}{6763.67} &\multicolumn{2}{c}{\textbf{6810.68}}\\

			&\multicolumn{1}{|c}{\textbf{SPEA-II}} &\multicolumn{2}{|c|}{5232.97} &\multicolumn{2}{c|}{5553.11} &\multicolumn{2}{c|}{4930.04} &\multicolumn{2}{c|}{11825.18} &\multicolumn{2}{c|}{11714.22} &\multicolumn{2}{c}{10559.02}\\

			&\multicolumn{1}{|c}{\textbf{MODA}} &\multicolumn{2}{|c|}{6448.75} &\multicolumn{2}{c|}{7179.98} &\multicolumn{2}{c|}{5249.45} &\multicolumn{2}{c|}{11547.61} &\multicolumn{2}{c|}{11543.43} &\multicolumn{2}{c}{11450.79}\\

			&\multicolumn{1}{|c}{\textbf{MSSA}} &\multicolumn{2}{|c|}{5587.69} &\multicolumn{2}{c|}{5401.46} &\multicolumn{2}{c|}{4276.64} &\multicolumn{2}{c|}{10842.81} &\multicolumn{2}{c|}{11172.36} &\multicolumn{2}{c}{13520.85}\\

			&\multicolumn{1}{|c}{\textbf{NSGA-II}} &\multicolumn{2}{|c|}{3739.69} &\multicolumn{2}{c|}{3827.96} &\multicolumn{2}{c|}{3615.98} &\multicolumn{2}{c|}{7441.22} &\multicolumn{2}{c|}{7700.99} &\multicolumn{2}{c}{7306.13}\\
			
			&\multicolumn{1}{|c}{\textbf{NSGA-II-KV}} &\multicolumn{2}{|c|}{\textbf{3557.07}} &\multicolumn{2}{c|}{\textbf{3585.88}} &\multicolumn{2}{c|}{\textbf{3509.03}} &\multicolumn{2}{c|}{6582.77} &\multicolumn{2}{c|}{\textbf{6729.54}} &\multicolumn{2}{c}{6828.00}\\\hline
			
		\end{tabular}
	\end{center}
\end{table*}

\par Moreover, the complexity of $K$-means algorithm is $\mathcal{O}(N\cdot M\cdot NP\cdot G_{max}\cdot I_1)$ \cite{DBLP:journals/access/StiawanSSIAAB21}, and the complexity of other comparison algorithms are the same as NSGA-II-KV \cite{DBLP:journals/tec/Jensen03a} \cite{2016Dragonfly} \cite{2017Salp}. The CPU running time of the NSGA-II-KV and other comparison algorithms are shown in Table \ref{Time-NSGA-II-KV}. It can be seen from the table that the running times of NSGA-II-KV do not increase compared to the conventional NSGA-II, and the gaps between NSGA-II-KV and other comparison algorithms excluding $K$-means are not very large. Although the $K$-means algorithm runs faster than other algorithms, it performs worse in the small size networks and $K$-means algorithm will damage $f_3$ according to the abovementioned statement. In addition, the proposed UAV-based charging and data collection is usually performed off-line. Thus, we may say that NSGA-II-KV obtains the overall best performance for coping with the UPAOP.
\begin{table*}[t]
	\scriptsize
	\begin{center}
		\caption{Numerical statistical results of CPU running times obtained by NSGA-II-KV and other comparison algorithms for solving UPAOP (s)}
		\label{Time-NSGA-II-KV}	
		\begin{tabular}{l|l|l|l|l|l|l|l|l|l|l|l|l}\hline
			\textbf{Algorithm}&\multicolumn{2}{c|}{\textbf {Scenario 1}}&\multicolumn{2}{c|}{\textbf {Scenario 2}}&\multicolumn{2}{c|}{\textbf {Scenario 3}}&\multicolumn{2}{c|}{\textbf {Scenario 4}}&\multicolumn{2}{c|}{\textbf {Scenario 5}}&\multicolumn{2}{c}{\textbf {Scenario 6}}\\\hline
			\textbf{$K$-means} &\multicolumn{2}{c|}{\textbf{0.01}} &\multicolumn{2}{c|}{\textbf{0.01}} &\multicolumn{2}{c|}{\textbf{0.01}} &\multicolumn{2}{c|}{\textbf{0.01}} &\multicolumn{2}{c|}{\textbf{0.01}} &\multicolumn{2}{c}{\textbf{0.01}}\\
			
			\textbf{SPEA-II} &\multicolumn{2}{c|}{53.32} &\multicolumn{2}{c|}{45.35} &\multicolumn{2}{c|}{45.60} &\multicolumn{2}{c|}{76.06} &\multicolumn{2}{c|}{75.06} &\multicolumn{2}{c}{76.60}\\
			
			\textbf{MODA} &\multicolumn{2}{c|}{36.66} &\multicolumn{2}{c|}{31.93} &\multicolumn{2}{c|}{31.33} &\multicolumn{2}{c|}{60.31} &\multicolumn{2}{c|}{56.14} &\multicolumn{2}{c}{58.96}\\
			
			\textbf{MSSA} &\multicolumn{2}{c|}{3.52} &\multicolumn{2}{c|}{3.39} &\multicolumn{2}{c|}{3.31} &\multicolumn{2}{c|}{10.50} &\multicolumn{2}{c|}{10.39} &\multicolumn{2}{c}{10.54}\\
			
			\textbf{NSGA-II} &\multicolumn{2}{c|}{27.22} &\multicolumn{2}{c|}{26.93} &\multicolumn{2}{c|}{27.05} &\multicolumn{2}{c|}{36.30} &\multicolumn{2}{c|}{36.12} &\multicolumn{2}{c}{35.98}\\
			\textbf{NSGA-II-KV} &\multicolumn{2}{c|}{26.87} &\multicolumn{2}{c|}{26.64} &\multicolumn{2}{c|}{28.08} &\multicolumn{2}{c|}{38.62} &\multicolumn{2}{c|}{37.23} &\multicolumn{2}{c}{36.24}\\\hline
			
		\end{tabular}
	\end{center}
\end{table*}
\begin{figure}[t]
	\scriptsize
	\centering{\includegraphics[width=3.5in]{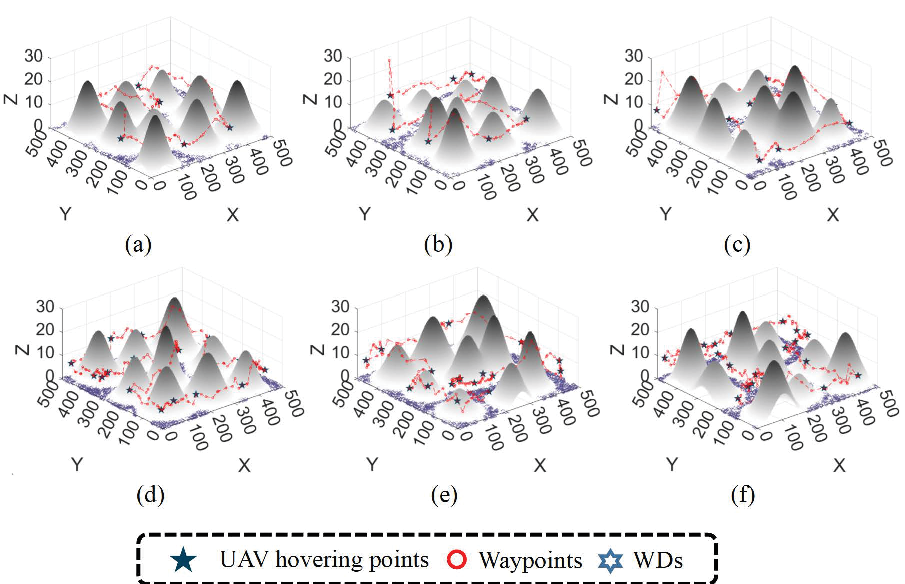}}
	\caption{3D trajectory optimization results obtained by PSO-NGDP for solving UTTOP. (a) Scenario 1. (b) Scenario 2. (c) Scenario 3. (d) Scenario 4. (e) Scenario 5. (f) Scenario 6.}
	\label{Trajectory}
\end{figure}

\subsection{Results of UTTOP}

\par
\begin{figure}[t]
	\centering{\includegraphics[width=3.5in]{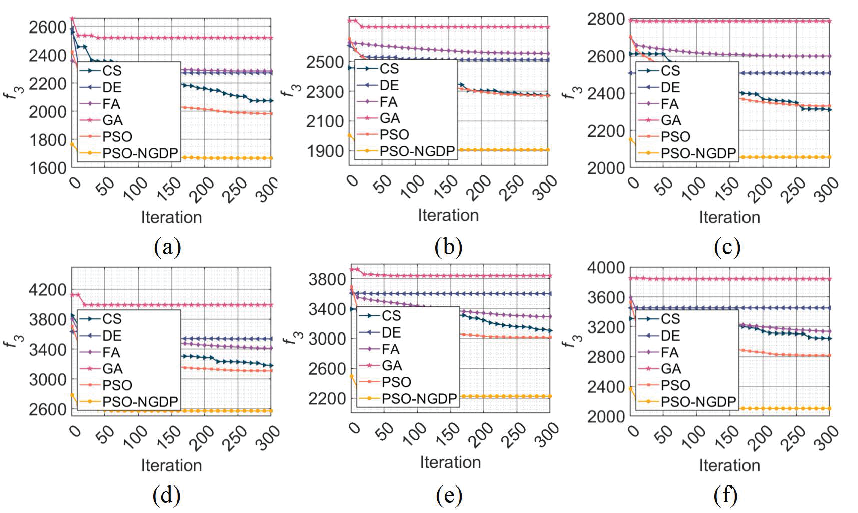}}
	\caption{Convergence rates obtained by PSO-NGDP and other comparison algorithms for solving UTTOP. (a) Scenario 1. (b) Scenario 2. (c) Scenario 3. (d) Scenario 4. (e) Scenario 5. (f) Scenario 6.}
	\label{convergence}
\end{figure}

\begin{figure}[t]
	\centering{\includegraphics[width=3.5in]{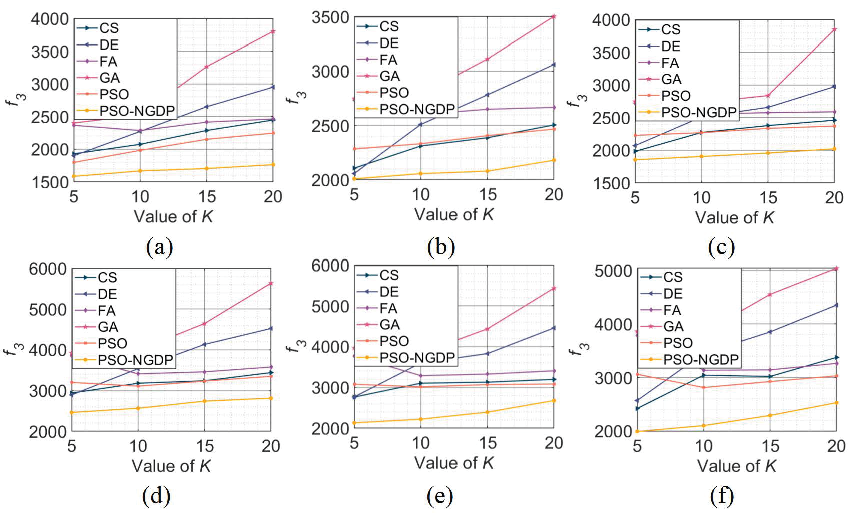}}
	\caption{Performance of different algorithms under different $K$. (a) Scenario 1. (b) Scenario 2. (c) Scenario 3. (d) Scenario 4. (e) Scenario 5. (f) Scenario 6.}
	\label{Different_K}
\end{figure}

 Figs. \ref{Trajectory}(a)-\ref{Trajectory}(f) show the 3D trajectory optimization results obtained by PSO-NGDP. As can be seen, UAV can avoid obstacles by using PSO-NGDP. Moreover, the convergences of different algorithms for Scenarios 1-6 are shown in Figs. \ref{convergence}(a)-\ref{convergence}(f), respectively. Specifically, we select cuckoo search (CS) \cite{7389387}, DE \cite{6919306}, firefly algorithm (FA) \cite{liang2021charging}, GA \cite{SQUIRES2022116464} and PSO \cite{5675669} as the comparison algorithms, and the complexities of these algorithms are the same as PSO-NGDP. It can be seen from Figs. \ref{convergence}(a)-\ref{convergence}(f), PSO-NGDP owns the fastest convergence speeds and the best results under each scenario, which means that the UAV consumes the least flight energy. Moreover, we test the results of different algorithms under different $K$ namely different number of waypoints which can be seen in Figs. \ref{Different_K}(a)-\ref{Different_K}(f). In addition, Table \ref{PerformancePSONGDP} shows the numerical results of PSO-NGDP and other comparison algorithms when $K=10$. It can be seen from the figures and the Tables \ref{PerformancePSONGDP} and \ref{Time-PSO-NGDP} that the proposed PSO-NGDP has better performance than other comparison algorithms while it takes more CPU running time. This is because the introduced improved factors of PSO-NGDP may need extra calculations and comparisons, which cause more CPU running time. As mentioned above, the considered method is usually performed off-line. Thus, we may say that the PSO-NGDP is the most effective approach for dealing with the UTTOP.

\begin{table*}[!t]
	\scriptsize
	\centering
	\caption{Numerical statistical results obtained by PSO-NGDP and other comparison algorithms for solving UTTOP when $K=10$ (m)}	
	\label{PerformancePSONGDP}
	\begin{center}
		\begin{tabular}{l|l|l|l|l|l|l|l|l|l|l|l|l}\hline
			&\multicolumn{2}{c|}{\textbf {Scenario 1}}&\multicolumn{2}{c|}{\textbf {Scenario 2}}&\multicolumn{2}{c|}{\textbf {Scenario 3}}&\multicolumn{2}{c|}{\textbf {Scenario 4}}&\multicolumn{2}{c|}{\textbf {Scenario 5}}&\multicolumn{2}{c}{\textbf {Scenario 6}}\\\hline			
			
			\textbf{CS} &\multicolumn{2}{c|}{2075.76} &\multicolumn{2}{c|}{2272.78} &\multicolumn{2}{c|}{2310.99} &\multicolumn{2}{c|}{3180.91} &\multicolumn{2}{c|}{3103.86} &\multicolumn{2}{c}{3043.19}\\
			
			\textbf{DE} &\multicolumn{2}{c|}{2273.40} &\multicolumn{2}{c|}{2510.73} &\multicolumn{2}{c|}{2509.00} &\multicolumn{2}{c|}{3539.03} &\multicolumn{2}{c|}{3599.27} &\multicolumn{2}{c}{3451.27}\\
			
			\textbf{FA} &\multicolumn{2}{c|}{2286.36} &\multicolumn{2}{c|}{2553.49} &\multicolumn{2}{c|}{2600.26} &\multicolumn{2}{c|}{3411.38} &\multicolumn{2}{c|}{3290.10} &\multicolumn{2}{c}{3136.02}\\
			
			\textbf{GA} &\multicolumn{2}{c|}{2522.50} &\multicolumn{2}{c|}{2729.61} &\multicolumn{2}{c|}{2785.51} &\multicolumn{2}{c|}{3991.65} &\multicolumn{2}{c|}{3842.25} &\multicolumn{2}{c}{3842.28}\\
			
			\textbf{PSO} &\multicolumn{2}{c|}{1983.00} &\multicolumn{2}{c|}{2268.83} &\multicolumn{2}{c|}{2331.35} &\multicolumn{2}{c|}{3110.30} &\multicolumn{2}{c|}{3016.08} &\multicolumn{2}{c}{2814.02}\\
			
			\textbf{PSO-NGDP} &\multicolumn{2}{c|}{\textbf{1667.08}} &\multicolumn{2}{c|}{\textbf{1904.53}} &\multicolumn{2}{c|}{\textbf{2055.85}} &\multicolumn{2}{c|}{\textbf{2569.43}} &\multicolumn{2}{c|}{\textbf{2223.44}} &\multicolumn{2}{c}{\textbf{2102.65}}\\\hline
			
		\end{tabular}
	\end{center}
\end{table*}
\begin{table*}[t]
	\scriptsize
	\caption{Numerical statistical results of CPU running times obtained by PSO-NGDP and other comparison algorithms for solving UTTOP when $K=10$ (s)}
	\label{Time-PSO-NGDP}
	\begin{center}
		\begin{tabular}{l|l|l|l|l|l|l|l|l|l|l|l|l}\hline
			&\multicolumn{2}{c|}{\textbf {Scenario 1}}&\multicolumn{2}{c|}{\textbf {Scenario 2}}&\multicolumn{2}{c|}{\textbf {Scenario 3}}&\multicolumn{2}{c|}{\textbf {Scenario 4}}&\multicolumn{2}{c|}{\textbf {Scenario 5}}&\multicolumn{2}{c}{\textbf {Scenario 6}}\\\hline			
			
			\textbf{CS} &\multicolumn{2}{c|}{12.73} &\multicolumn{2}{c|}{8.35} &\multicolumn{2}{c|}{14.36} &\multicolumn{2}{c|}{19.51} &\multicolumn{2}{c|}{20.76} &\multicolumn{2}{c}{16.45}\\
			
			\textbf{DE} &\multicolumn{2}{c|}{20.21} &\multicolumn{2}{c|}{9.13} &\multicolumn{2}{c|}{22.10} &\multicolumn{2}{c|}{28.07} &\multicolumn{2}{c|}{28.83} &\multicolumn{2}{c}{16.53}\\
			
			\textbf{FA} &\multicolumn{2}{c|}{9.43} &\multicolumn{2}{c|}{8.93} &\multicolumn{2}{c|}{9.20} &\multicolumn{2}{c|}{15.39} &\multicolumn{2}{c|}{15.47} &\multicolumn{2}{c}{15.22}\\
			
			\textbf{GA} &\multicolumn{2}{c|}{6.77} &\multicolumn{2}{c|}{6.60} &\multicolumn{2}{c|}{6.89} &\multicolumn{2}{c|}{12.56} &\multicolumn{2}{c|}{13.28} &\multicolumn{2}{c}{12.51}\\
			
			\textbf{PSO} &\multicolumn{2}{c|}{\textbf{6.21}} &\multicolumn{2}{c|}{\textbf{6.12}} &\multicolumn{2}{c|}{\textbf{6.33}} &\multicolumn{2}{c|}{\textbf{11.71}} &\multicolumn{2}{c|}{\textbf{11.69}} &\multicolumn{2}{c}{\textbf{11.53}}\\
			
			\textbf{PSO-NGDP} &\multicolumn{2}{c|}{38.92} &\multicolumn{2}{c|}{38.72} &\multicolumn{2}{c|}{38.56} &\multicolumn{2}{c|}{111.16} &\multicolumn{2}{c|}{110.13} &\multicolumn{2}{c}{110.97}\\\hline
			
		\end{tabular}
	\end{center}
\end{table*}
\subsection{Stability tests of NSGA-II-KV and PSO-NGDP}
\par Due to the nature of the multi-objective optimization algorithms, they can obtain a set of Pareto solutions, which means the decision makers need to select one solution according to the real scenario. For example, there are some circumstances that the UAVs are applied for post-disaster constructions. Under these circumstances, the UAVs require to cover as more WDs as possible. Moreover, in some regular scenarios, the UAVs need to increase the time efficiency so that increasing the flying distance. Thus, we design two strategies to be selected for decision makers, which are covered WD maximization strategy (CWMS) and time efficiency maximization strategy (TEMS). Specifically, we select Scenarios 1 and 4 to test the stability of the proposed NSGA-II-KV for small and large size networks, respectively.

\par Table \ref{numerical results of UPAOP by CWMS} and Table \ref{numerical results of UPAOP by TEMS} show the numerical results of CWMS and TEMS obtained by NSGA-II-KV and other comparison algorithms for solving UPAOP, respectively. Note that ``Mean'', ``Std.'', ``Maximum'' and ``Minimum'' represent the mean value, standard deviation, maximum value and minimum value of $30$ independent trials. As can be seen, for both of CWMS and TEMS, the proposed NSGA-II-KV obtains the best performance in Scenario 1, while it achieves the best performance on $f_1$ and the second best performance on $f_2$ in Scenario 4. Moreover, the gaps on $f_2$ are not very large. Thus, we can say the proposed NSGA-II-KV is stable. Moreover, the stability test results for solving UTTOP are shown in Table \ref{Stability test of UTTOP}, where the proposed PSO-NGDP obtains the best stability.
\begin{table*}[h]
	\scriptsize   
	\begin{center}
		\caption{Numerical statistical results of CWMS obtained by NSGA-II-KV and other comparison algorithms for solving UPAOP}
		
		{\begin{tabular}{llllllllll}\hline
				
				&&{\textbf{US}} &\textbf{RS} &\textbf{$K$-means}&\textbf{SPEA-II}&\textbf {MODA} &\textbf{MSSA} &\textbf{NSGA-II} &\textbf{NSGA-II-KV}\\\hline
				
				\multicolumn{10}{c}{\textbf{Scenario 1}}\\\hline
				
				\multirow{4}{*}{\textbf {$f_1$}}  
				&Mean& 5.00& 15.03& 84.83&91.63& 84.83&105.47&128.90&\textbf{138.80}\\

				&Std.&\textbf{0.00}&8.50&4.77&9.30&4.77&6.75&11.15&5.18\\				
				
				&Maximum&5.00&36.00&94.00&107.00&96.00&119.00&148.00&\textbf{154.00}\\				
				
				&Minimum&5.00&1.00&76.00&77.00&74.00&93.00&97.00&\textbf{132.00}\\\hline
				
				\multirow{4}{*}{\textbf {$f_2$}}  
				&Mean& 8026.63& 8319.07& 6973.69& 5424.26& 6973.69&5487.22&3886.75&\textbf{3733.87}\\

				&Std.&\textbf{0.00}&840.98&794.39&285.53&794.39&743.21&282.76&178.74\\				
				
				&Maximum&8026.63&9437.72&8605.73&6005.85&8605.73&7271.63&5085.93&\textbf{4307.82}\\				
				
				&Minimum&8026.63&6419.93&4833.54&4845.13&4833.54&4149.32&3670.07&\textbf{3405.89}\\\hline

				\multicolumn{10}{c}{\textbf{Scenario 4}}\\\hline
				
				\multirow{4}{*}{\textbf {$f_1$}}  
				&Mean& 63.00&40.47& 396.50& 221.50&316.20&346.13&339.93&\textbf{441.80}\\

				&Std.&\textbf{0.00}&14.05&16.25&28.23&10.63&10.72&32.96&9.68\\				
				
				&Maximum&63.00&62.00&446.00&281.00&340.00&366.00&406.00&\textbf{457.00}\\				
				
				&Minimum&63.00&4.00&368.00&161.00&301.00&319.00&281.00&\textbf{417.00}\\\hline
				
				\multirow{4}{*}{\textbf {$f_2$}}  
				&Mean&14558.10&16670.52&\textbf{6471.80}&11708.32&12101.86&12061.98&7723.70&6986.95\\

				&Std.&\textbf{0.00}&1133.98&133.52&817.74&1186.59&989.17&503.03&343.22\\				
				
				&Maximum&14558.10&19435.20&\textbf{6994.09}&13565.85&15672.24&14550.58&9307.67&7738.13\\				
				
				&Minimum&14558.10&14146.21&\textbf{6265.65}&10210.21&10489.36&10387.15&7077.60&6551.64\\\hline

		\end{tabular}}
		\label{numerical results of UPAOP by CWMS}
	\end{center}
\end{table*}

\begin{table*}[h]  
	\scriptsize
	\begin{center}
		\caption{Numerical statistical results of TEMS obtained by NSGA-II-KV and other comparison algorithms for solving UPAOP}
		{\begin{tabular}{llllllllll}\hline
				
				&&{\textbf{US}} &\textbf{RS} &\textbf{$K$-means}&\textbf{SPEA-II}&\textbf {MODA} &\textbf{MSSA} &\textbf{NSGA-II} &\textbf{NSGA-II-KV}\\\hline

				\multicolumn{10}{c}{\textbf{Scenario 1}}\\\hline
				
				\multirow{4}{*}{\textbf {$f_1$}}  
				&Mean&5.00&15.03&84.83&78.60&51.40&86.70&110.63&\textbf{114.83}	\\

				&Std.&\textbf{0.00}&8.50&4.77&10.21&10.81&15.42&13.59&12.84\\				
				
				&Maximum&5.00&36.00&94.00&98.00&74.00&115.00&\textbf{137.00}&131.00\\				
				
				&Minimum&5.00&1.00&76.00&57.00&36.00&51.00&80.00&\textbf{86.00}\\\hline
				
				\multirow{4}{*}{\textbf {$f_2$}}  
				&Mean&8026.63&8319.07&6973.69&5204.40&5182.04&4872.61&3678.35&\textbf{3448.19}\\

				&Std.&\textbf{0.00}&840.98&794.39&265.86&565.75&334.51&256.54&88.62\\				
				
				&Maximum&8026.63&9437.72&8605.73&5617.48&5819.84&5435.80&4870.80&\textbf{3623.84}\\				
				
				&Minimum&8026.63&6419.93&4833.54&4661.42&3930.93&3994.79&3500.81&\textbf{3292.32}\\\hline

				\multicolumn{10}{c}{\textbf{Scenario 4}}\\\hline
				
				\multirow{4}{*}{\textbf {$f_1$}}  
				&Mean&63.00&40.47&396.50&213.70&245.93&283.50&321.96&\textbf{428.40}\\					
				
				&Std.&\textbf{0.00}&14.05&16.25&28.49&39.29&30.55&34.87&13.69\\				
				
				&Maximum&63.00&62.00&446.00&270.00&298.00&352.00&388.00&\textbf{447.00}\\				
				
				&Minimum&63.00&4.00&368.00&155.00&152.00&214.00&252.00&\textbf{398.00}\\\hline
				
				\multirow{4}{*}{\textbf {$f_2$}}  
				&Mean&14558.10&16670.52&\textbf{6471.80}
				&11610.69&10432.97&10310.23&7408.88&6735.41\\

				&Std.&\textbf{0.00}&1133.98&133.52&807.42&542.46&409.85&391.71&188.09\\				
				
				&Maximum&14558.10&19435.20&\textbf{6994.09}&13492.73&11523.84&11011.11&9155.38&7203.73\\				
				
				&Minimum&14558.10&14146.21&\textbf{6265.65}&10163.28&9466.55&9449.71&6995.44&6379.17\\\hline

		\end{tabular}}
		\label{numerical results of UPAOP by TEMS}
	\end{center}
\end{table*}
\begin{table*}[h]
	\scriptsize
	\begin{center}
		\caption{Numerical statistical results of UTTOP obtained by PSO-NGDP and other comparison algorithms}
		{\begin{tabular}{l|l|l|l|l|l|l|l|l|l|l|l|l|l}\hline
				&&\multicolumn{6}{c|}{\textbf {Scenario 1}}&\multicolumn{6}{c}{\textbf {Scenario 4}}\\\hline
				&&{\textbf{CS}} &\textbf{DE} &\textbf {FA} &\textbf{GA} &\textbf{PSO} &\textbf{PSO-NGDP}&{\textbf{CS}}&\textbf{DE} &\textbf {FA} &\textbf{GA} &\textbf{PSO} &\textbf{PSO-NGDP}\\\hline
				\multirow{4}{*}{\textbf {$f_3$}}
				
				&Mean&2085.23&2288.96&2286.93&2536.89&1980.18&\textbf{1657.25}&3221.90&3629.58&3399.09&4017.09&3123.08&\textbf{2548.36}\\	
				
				&Std.&16.44&22.23&\textbf{9.00}&63.61&19.71&9.70&20.30&21.99&\textbf{11.96}&89.39&26.27&13.51\\				
				
				&Maximum&2120.66&2334.87&2300.47&2656.27&2031.81&\textbf{1678.26}&3277.97&3683.07&3423.98&4162.00&3163.08&\textbf{2572.94}\\ 
				
				&Minimum&2055.31&2243.18&2258.56&2402.78&1941.79&\textbf{1636.37}&3169.11&3590.23&3380.67&3757.53&3075.90&\textbf{2520.21}\\\hline
		\end{tabular}}
		\label{Stability test of UTTOP}
	\end{center}
\end{table*}
\section{Discussion}
\label{Discussion}
\subsection{The influence of propulsion power of UAV} 
\par The propulsion power is only affected by the flying velocity according to Eq. (\ref{UAV-2D-Power}). It can be seen from Fig. \ref{charging Distance and PV}(a) which can be drawn according to Eq. (\ref{UAV-2D-Power}), when UAV velocity $V\approx 10.2$, the propulsion power is the minimum power. The purpose of this work is to improve the energy utilization efficiency, and we wonder to save the energy consumption of UAV. Thus, the velocity of the UAV can be chosen as a constant which can obtain the minimum propulsion power. Moreover, if the scenario has the requirements about task completing time, the UAV needs to fly faster, which leads to the increasing energy consumption. Under this circumstance, the UAV with better performance will be chosen.
\subsection{The rationality of the maximum and minimum altitudes of UAV}
\par
\begin{figure}[htb]
	\centering{\includegraphics[width=3.5in]{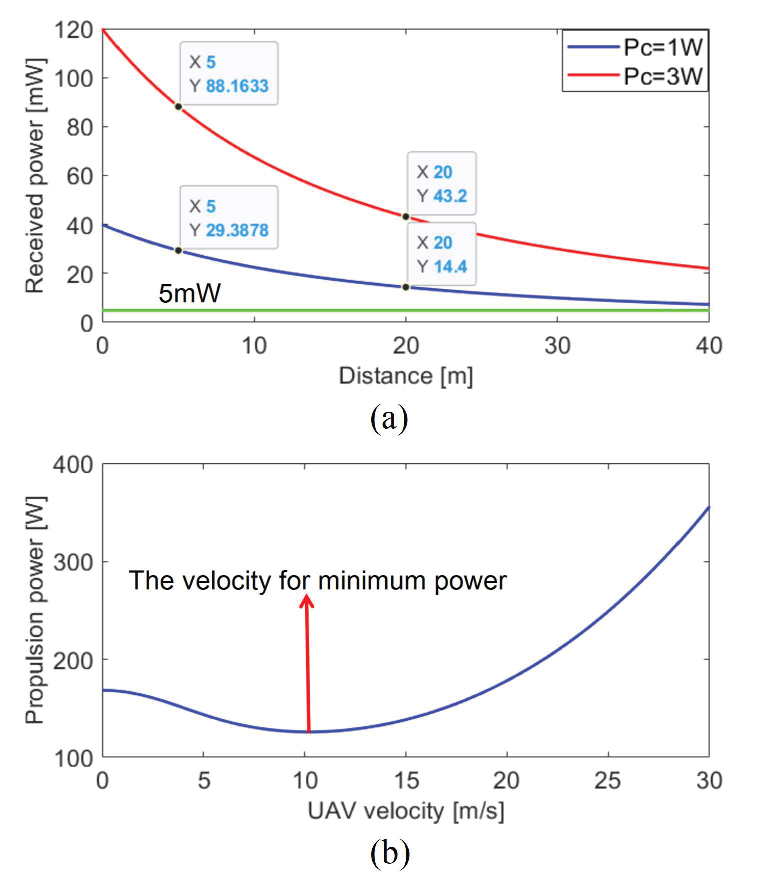}}
	\caption{(a) Typical plots for UAV propulsion power consumption versus velocity. (b) Relationship between the charging distance and the received power.}
	\label{charging Distance and PV}
\end{figure}

 As above mentioned, the charging distance threshold is selected as the ultimate distance threshold in this paper. In other words, if the UAV can achieve charging, it can also achieve data collection. Thus, we only need to consider the influence of the minimum and maximum altitudes for charging. Specifically, we can set the maximum power of UAV to be 3 W, which is the rated power of a widely-used transmitter, i.e., TX91501 transmitter of the Powercast device \cite{sun2021trading}. Then, we follow the charging parameters of \cite{liang2021charging}. Also, the received power threshold is set as $5$ mW since it can cover the most practical scenarios \cite{DBLP:journals/cn/LiSWLLKL22}. It can be seen from Fig. \ref{charging Distance and PV}(b), the received power obtained by WDs are still larger than the minimum received power threshold which is $5$ mW, in which $1$ W is the minimum transmission power of UAV which can be achieved by power splitting technology \cite{DBLP:journals/twc/LiCWWL22}. Thus, the charging distances which are $5$ m and $20$ m in this work are feasible settings.

\par Moreover, note that the setting of $20$ m is only for charging and data collection. However, the maximum flying altitude $Z_{max}$ is set as $30$ m, and the setting of $30$ m is only for avoiding the obstacles instead of charging and data collection, since the altitudes of obstacles may be larger than $20$ m. Such a value can be changed by different shapes of obstacles. With the increasing of the obstacle altitude, it can set a larger value to avoid obstacles.
\section{Conclusion}
\label{Conclusion}

\par In this paper, the joint-UAV power and 3D trajectory optimization in WPCN for improving the energy utilization efficiency is investigated. First, we consider that a UAV needs to cover as more WDs as possible while improves time efficiency and reduces the energy consumption. Specifically, we formulate a JUPTTOP to jointly increase the total number of the covered WDs, increase the time efficiency, and reduce the motion energy consumption of UAV. Then, due to the difficulties and complexities of JUPTTOP, we divide it into UPAOP and UTTOP, respectively, and propose an NSGA-II-KV and a PSO-NGDP to solve them. Simulations are conducted to evaluate the performances of the two proposed algorithms, and the results demonstrate that the proposed algorithms achieve the better performances for UPAOP and UTTOP than other comparison algorithms, respectively. In the future work, we may consider further optimization of the algorithm to avoid unknown obstacles in near real time.

\bibliographystyle{ieeetr}
\bibliography{ref-MyUAV}

\end{document}